\documentclass[preprint2]{aastex}

\begin{document}
\title{High-Resolution Spectroscopy of FUors }
\author{G. H. Herbig }
\affil{Institute for Astronomy, University of Hawaii,\\  2680 Woodlawn Drive,
Honolulu, Hawaii 96822, U.S.A. }
\author{P. P. Petrov }
\affil{Crimean Astrophysical Observatory,  p/o Nauchny, Crimea 98409, Ukraine\\ and\\ Isaac Newton Institute
of Chile, Crimean Branch }

\and

\author{R. Duemmler\footnote{Present address: Frankenallee 201, D-60326 Frankfurt/Main, Germany }
 }
\affil{Astronomy Division, University of Oulu, P.O. Box 3000, FIN--90014, Finland}
\singlespace

\begin{abstract}
	High-resolution spectroscopy was obtained of the FUors FU Ori and
V1057 Cyg between 1995 and 2002 with the SOFIN spectrograph at
NOT and with HIRES at Keck I. During these years FU Ori remained about 1 mag.\
(in B) below its 1938--39 maximum brightness, but V1057 Cyg (B $\approx$ 10.5
at peak in 1970--71) faded from about 13.5 to 14.9 and then recovered slightly.
Their photospheric spectra resemble that of a rotationally broadened, slightly
veiled supergiant of about type G0 Ib, with $v_{\rm eq}\, \sin i = 70$ km s$^{-1}$
for FU Ori, and 55 km s$^{-1}$ for V1057 Cyg.  As V1057 Cyg faded, P Cyg
structure
in H$\alpha$ and the IR \ion{Ca}{2} lines strengthened, and a complex
shortward-displaced shell spectrum of low-excitation lines of the neutral
metals (including \ion{Li}{1} and \ion{Rb}{1}) increased in strength, disappeared in
1999, and reappeared in 2001. Several SOFIN runs extended over a number
of successive nights so that a search for rapid and cyclic changes in the
spectra was possible.  These spectra show rapid night-to-night changes in
the wind structure of FU Ori at H$\alpha$, including clear evidence of
sporadic infall.  The equivalent width of the P Cyg
absorption varied cyclically with a period of 14.8 days, with
phase stability  maintained over 3 seasons.  This is believed to be the
rotation period of FU Ori.  The internal structure of its photospheric lines
also varies cyclically, but with a period of 3.54 days.  A similar variation
may be present in V1057 Cyg, but the data are much noisier and that result
uncertain.  As V1057 Cyg has faded and the continuum level fallen, the
emission lines of a pre-existing low-excitation chromosphere have emerged.
Therefore we believe that the `line doubling' in V1057 Cyg is produced by
these central emission cores in the absorption lines, not by orbital motion
in an inclined Keplerian disk. No dependence of $v_{\rm eq}\, \sin i$ on
wavelength or excitation potential was detected in either FU Ori or V1057 Cyg,
again contrary to expectation for
a self-luminous accretion disk.  It was found also that certain critical lines
in the near infrared are not accounted for by synthetic disk spectra.  It is
concluded that a rapidly rotating star near the edge of stability, as proposed
by Larson (1980), can better account for these observations.  The possibility
is also considered that FUor eruptions are not a property of ordinary T
Tauri stars, but may be confined to a special sub-species of rapidly rotating
pre-main sequence stars having powerful quasi-permanent winds.
\end{abstract}


\section{ Introduction}
\normalsize

	The variable star now known as FU Ori was originally believed
to be a slow nova because of its leisurely rise from m$_{pg}$ = 16 to
10 over an interval of about a year in 1937--39.  There were some misgivings
at the time about that classification: the spectrum was quite unlike that of
an ordinary nova, and for a nova the star was most unusually located in the
dark cloud Barnard 35, itself in the OB association surrounding $\lambda$
Ori.  It is now realized that FU Ori is no nova, but represents another
phenomenon altogether. It is the prototype of a small class of pre-main
sequence objects, named `FUors' by Ambartsumian, that have received
an increasing amount of attention over the past three decades.

	Three additional stars, and possibly 2 more, now collectively define
the FUor class on the grounds of a well-documented major rise
in brightness, association with a molecular cloud, and a spectrum like
that of FU Ori.  These `classical FUors' are V1057 Cyg, V1515 Cyg, V1735
Cyg, probably V346 Nor \citep{rei90}, and possibly a star (CB34V = V1184 Tau)
discussed most recently by \citet{alv97}.  A number of
other pre-main sequence stars have been proposed for membership, usually on
the grounds of spectroscopic resemblance plus an infrared excess, but none have been
observed to brighten up and remain so for years, as did the classical FUors
mentioned above.
  
	 The time seems appropriate for a reexamination of the observational
situation for two reasons.  First, the other classical FUors remain not far
from their maximum brightness, but V1057 Cyg has faded about 4 mag.\ (in B)
since 1971.

	 Second, high-resolution spectroscopy has now become feasible
for all FUors over a wider wavelength range (3500--9000 \AA\ ) than was
heretofore possible, and in the case of V1057 Cyg almost on an annual basis.
The observations to be discussed here have been obtained by Petrov and
collaborators with the SOFIN echelle spectrograph \citep{tuo99}
of the Nordic Optical
telescope at La Palma\footnote { The Nordic Optical Telescope is
operated on the island of La Palma jointly by Denmark, Finland, Iceland,
Norway and Sweden, in the Spanish Observatorio del Roque de los Muchachos of
the Instituto de Astrofisica de Canarias. } 
 between 1996 and 2002 at a resolution of about 13
km s$^{-1}$; and by Herbig with the HIRES echelle at the Keck I telescope
on Mauna Kea\footnote {The W. M. Keck Observatory is operated as a
scientific partnership among the California Institute of Technology, the
University of California, and the National Aeronautics and Space
Administration.  The Observatory was made possible by the generous financial
support of the W. M. Keck Foundation.}
since 1996 at a resolution of about 7 km s$^{-1}$.
Several
of the SOFIN runs extended over a number of successive nights, offering the
opportunity to search for rapid changes in the spectra. \footnote {The dates
of the SOFIN observations are not tabulated here, but can be retrieved from
the listing in Table 9, this paper.} Some of the SOFIN
material has already been discussed by \citet{pet98} and by \citet{laa00}.
We here expand upon those results.

	It is now accepted that FUors represent an interesting phenomenon of
early stellar evolution, but it is uncertain how universal it is, and there
is disagreement on what is responsible for the outbursts.  Hypotheses as to
the latter fall into two classes: Hartmann, Kenyon, and their colleagues have
proposed that the flare-up is a phenomenon not of the pre-outburst star
itself, but is the result of a major increase in the surface brightness of the
circumstellar accretion disk. Since the idea was first put forward by
\citet[][hereafter HK]{har85}, it has been elaborated extensively 
\citet[and reviewed:][]{har96}.  Building upon that proposition, theories of
instabilities intrinsic to such an accretion disk have been examined by
\citet{cla90}, by \citet{kle99} and by  \citet[][which contains
references to earlier papers]{bel99}.

	The opposite hypothesis is that the star itself is responsible
for the FUor flare-up. The absorption lines in the classical FUors are
broad; if they are the result of axial rotation in a spherical limb-darkened
star, fits to the optical-region line profiles (\S\ 4.2, below) yield values
of $v_{\rm eq}\, \sin i$ up to about 70 km s$^{-1}$ in the case of FU Ori.
A periodic modulation of the line structure of FU Ori (described in
\S\ 4.4) leads to a
radius of the order of 20 sin $i$ R$_{\odot}$.  Given
those parameters, the condition that FU Ori be rotationally stable
(in the sense that the centrifugal acceleration at the equator of a
oblate rotator equals the gravitational [\citealt{por96}]) is
(M/M$_{\odot}$)\, sin $i$ $>$ 0.79.  Clearly, a not-unreasonable value
of sin $i$ would require a substantial stellar mass to ensure the stability
of such a rapid rotator.

  	An examination by \citet{lar80} of the consequences of such very
rapid rotation suggested that bar-like deformations would develop that
could produce
heating of the outer layers of the star, thus accounting for the flare-up
and mass loss.  One would think that such instabilities might produce
detectable photometric variation with the rotation period.  The only
search for such variations is that reported for FU Ori by
\citet{ken00}.
They found only `random' fluctuations of amplitude 3-4\% on time scales
of 1 day or less, but it would be worthwhile to repeat such observations with
better time coverage.  Later (\S\ 4.4) we describe the results of a
search for cyclic variations in radial velocity and line structure
in both FU Ori and V1057 Cyg.

	 Little more has been heard of the rapid-rotator hypothesis,
perhaps because of the appeal of the disk-instability idea and the volume of
publication that it has engendered. But on earlier occasions we have
pointed out some difficulties and our reservations about the HK proposal
\citep{her89,pet92}.  The theory was subsequently modified
to explain one of those concerns \citet{bel95}. 
However the new results to be described here call into question the
original observational justification for the HK hypothesis, on which
that theory is based.  In the following sections we examine
three issues that have been regarded as crucial support for that
hypothesis:

	(a) The `doubling' of certain absorption lines is evidence of
Keplerian motion in an inclined disk  (\S\ 2.4);
 
        (b) A dependence on $v_{\rm eq}\, \sin i$ on wavelength or excitation
potential in the optical region up to 9000 \AA\ is evidence of the
decline in orbital velocity with radius expected under the disk hypothesis
(\S\ 4.2); and

        (c) The theoretical disk spectrum, as a composite of annuli of
temperature and surface brightness that decline with distance from the
center, fits the observed spectra of FUors (\S\ 4.1).  

 	 When V1057 Cyg was bright, it was noticed that the spectral type
became later with increasing wavelength across the optical region.  The
effect became striking when observations were extended to the near-IR
\citep{mou78,eli78} and H$_{2}$O and CO bands were found in the
spectrum
of an ostensible F- or G-type star.  No evidence of radial velocity variation
with time or with wavelength has been reported, so the spectrum apparently
originates in a single object.  The disk hypothesis does have the persuasive
advantage of explaining, although not in detail (\S\ 4.1), this apparently
composite nature of FUor spectra as the falloff of temperature with increasing
radius in an accretion disk.

        But it is now realized that the simple presence of the 2.3-$\mu$m
CO bands in a G supergiant is not that unusual \citep{wal97}.
They appear at about G0 Ib and are prominent by G8 Ib, although in FU Ori and
V1057 Cyg they are as strong as in early M-type. It would be interesting to
investigate if such an effect could be simulated in an extended envelope
around a rapidly rotating single star.

	We are not necessarily committed to the concept of an unstable rapid
rotator as a final solution to the FUor phenomenon.  But for lack of a
more persuasive alternative, in what follows we examine the observational
information in terms of that hypothesis.  First we describe the spectrum of
V1057 Cyg during its 1996--2002
fading, the period for which we have detailed high-resolution coverage (\S\ 2),
followed by a discussion of its light curve (\S\ 3), and then
we review the spectroscopic properties of FUors in general (\S\ 4).

\section{ The Spectrum of V1057 Cyg 1996--2002}

	Figure 1 shows the B/pg light curve of V1057 Cyg.  Most
of the high-resolution spectra of V1057 Cyg discussed in the literature were
obtained in the 1980's, when the star was descending to a plateau of
brightness between about B = 13.0 and 13.3, which extended from
1985 to about 1994.  Between 1994 and 1995 it began to fade again, and
reached a minimum near 14.9 in 1999, since which it has 
recovered slightly.   Briefly, following \citet{pet98} and
\citet{laa00}, this is what happened in the spectrum during the
post-1994 minimum with respect to the 1980s:

\noindent
\hspace*{0.2in} the photospheric lines became much shallower, some with
    pronounced emission cores;\\
\hspace*{0.2in} the P Cyg absorptions in H$\alpha$, infrared \ion{Ca}{2} and
    other lines became stronger;\\ 
\hspace*{0.2in} the shortward-shifted shell components in the low excitation
lines increased in strength \\
\hspace*{0.4in} (1996-97), then essentially disappeared (1999),
but reappeared in 2001;  \\
\hspace*{0.2in} TiO bands in the expanding shell appeared for the first
 time, and \\
\hspace*{0.2in} a number of emission lines of low excitation appeared in
 the spectrum.\\

\subsection{ Photospheric spectrum}  

	Simple inspection of the absorption spectrum of V1057 Cyg
shows that it resembles that of a rotationally broadened early-G
supergiant. To make this quantitative, the excitation temperature and
gravity were determined as follows.  Equivalent widths (hereafter
abbreviated EW)  were measured for 30 of the least blended photospheric
lines of \ion{Fe}{1} in the region
4950-8000 \AA\  (shell components, which could distort the
measurements, were almost absent in this spectral region in 1998--1999). 
The curve of growth gives T$_{exc}$ = 5300 $\pm$ 300 K. The single
measurable line of \ion{Fe}{2}, $\lambda$ 5991, lies on this curve at a position
corresponding to log n$_{e}$ = 12.  These are indeed values expected for
a G1 supergiant.   Furthermore, a comparison of V1057 Cyg with templates
of 41 Cyg (F5 II), $\beta$ Aqr (G0 Ib), 9 Peg (G5 Ib) and 40 Peg (G8 II)
show that the line ratios correspond to F7-G3 I-II.

	The photospheric absorption lines in August 1997 appeared to be
shallower than in the spectra found in the literature.  Many of the lower-
excitation lines appear double, as the result of what is clearly an
emission component at the bottom of the absorption. With respect to
spectra of the 1980's, this line `doubling,'
measured as a velocity separation between the two absorption minima, had
increased by October 1996, and even more by August 1997, as the result of what is clearly an emission
component appearing at the bottom of the absorption line (see \S\ 2.4). 
However, the line width of the absorptions near the continuum level remains
the same as in the spectra taken,  e.g., in 1983--84 by
\citet*{ken88}  and in 1992 by \citet{har95}. That is, the overall line width
has not changed, but in those where the central reversal has become more
prominent the line depth has been reduced.   In this way many
photospheric lines appear significantly shallower than in the accretion
disk model of \citet{ken88}, which was designed to fit the
spectrum of V1057 Cyg in 1985 (Fig.\ 13).

	However, there is another effect as well: in the red, 
higher-excitation lines where no central emission is expected are also
shallower than in the spun-up standards, as if a veiling continuum is
superposed upon the absorption spectrum.  Veiling factors of 0.3 to 0.5 are
indicated. 
 
\subsection{ Wind/P Cyg Features}

 	Table 1 gives the parameters of the P Cyg structure
at H$\alpha$
as measured on all spectra either published or our own.  One sees that the
mass-loss in the wind of V1057 Cyg, as evidenced by the strength of
the P Cyg absorption component at H$\alpha$, began to increase about 1984-85
\citep{laa00}, near the beginning of the photometric
plateau.  It is tempting to speculate that the two may be connected. 
Earlier, the P Cyg absorption of H$\alpha$ was not saturated
and its structure varied considerably at all velocities
\citep*[][and Table 1]{cro87}.  During the post-plateau
minimum (i.e., after 1994--95) 
the line remained strong, but the high velocity wing ($-200$ to $-400$ km
s$^{-1}$) varied in depth from year to year while the low velocity
section ($-60$ to $-120$ km s$^{-1}$) always remained deep (Figs. 7, 24).

 	The only known observations of H$\alpha$ during the plateau decade
are those obtained in 1988 by \citet{wlt92} and by Budge (unpublished). 
Those entries in Table 1 have been measured from their profiles.  The EW
of the P Cyg absorption component in that year was the largest that
has been reported.  No other observations of H$\alpha$ are known to have been
made during 1986-95, but \citet{rus01} measured the structure
of H$\beta$ and H$\gamma$ between 1978 and 1990 (photographically and at a
resolution of about 1000).  His data indicate that the EW of H$\beta$
declined from 1978 to 1985 but had increased again by 1987
(no observation in 1986), continued to increase thereafter and by 1990
was the largest he had observed.

	The lack of adequate spectroscopic coverage of V1057 Cyg during the
1986-95 decade is regrettable.  

	Figure 2 shows the P Cyg profiles of the H$\alpha$,
H$\beta$, \ion{Na}{1} and \ion{Ca}{2} $\lambda$8542 lines on the 1997
August SOFIN spectrogram. 
(Throughout we denote the measured (heliocentric) velocity of a features
as $v$ and the stellar velocity as $v_{*}$, so velocity in the star's
rest frame  is $RV$ = $v$ -- $v_{*}$.)
The H$\alpha$ profile is similar to that of FU Ori, where the mass-loss
rate was estimated to be an order of magnitude larger than in V1057 Cyg
in 1985 \citep{cro87}.

\subsection{ Shell Features}
	
	The low excitation ($<$1 eV) photospheric absorption lines of V1057
Cyg have almost always been flanked shortward by narrow, often complex `shell'
features. They are prominent in the strong low-excitation lines of the neutral
metals at shorter wavelengths.  They are probably due to condensations in the
expanding wind passing in front of the star.  In 1996, by which time the star
had faded by about 1 mag.\ (in B) from plateau brightness, both the P Cyg
absorption at the Balmer lines and the shell lines had become stronger, to such
a degree that the shell also became detectable
at many lines in the red. 

	Figure 3 shows some shell line profiles at the 13
km s$^{-1}$ SOFIN resolution. The underlying photospheric spectrum has
been subtracted, as described later (\S\ 2.5).  The strong lines of higher
excitation potential (hereafter EP), as \ion{Mg}{1} $\lambda$5183 and
 \ion{Fe}{2}
$\lambda$5316, also contain shortward shell components but at more negative
velocities, about $-120$ km s$^{-1}$. These absorptions are much broader than
those at the low excitation lines, and so represent an intermediate
case between shell and wind. 

	Another indicator of a cool expanding shell are the numerous TiO 
bands (Fig.\ 4). All the TiO bands were blue-shifted to RV = $-40$ to 
$-70$ km s$^{-1}$; i.e. to about the same velocity as the shell components
of low excitation lines. These shell lines and the TiO
bands in the red were strongest in 1996, weaker in 1997, and absent in 1998,
1999 and 2000.  By 2001 the shell lines and TiO bands had reappeared, but
with somewhat broader profiles and larger expansion velocities (see Fig.\ 3,
bottom).

	The complexity of the shell structure is more apparent at
HIRES resolution, where at least five separate components are seen. 
The shell was so prominent in 1996--97 at low-level lines of neutral metals
(\ion{Al}{1}, \ion{Fe}{1}, \ion{Ti}{1}, \ion{Cr}{1}, \ion{Mn}{1}), 
as well as in \ion{Ba}{2}, \ion{Li}{1},
and \ion{Rb}{1} as to confuse the shortward wings of the underlying stellar
features.  The 4990--5020 \AA\ region, containing a number of prominent
\ion{Ti}{1} lines, is shown in Figure 5.  At the top of the
figure is the same section in the G5 Ib star HD 190113, as broadened by
$v_{\rm eq}\, \sin i$ = 55 km s$^{-1}$, to represent the underlying photospheric
spectrum.

	At that time (1997 Aug. 12-13), the shell lines were at velocities
of $-128$, $-107$, $-89$,
$-78$, and $-61$ km s$^{-1}$; the velocity of the star itself is near $-16$
km s$^{-1}$.  A number of these complex shell lines
have been decomposed in the following way.  Each component was
represented as a pure absorption line of depth e$^{-\tau}$, where
$\tau$ was gaussian (of the form exp$(-0.5(\delta v/\sigma)^{2}$)) whose
velocity, depth, and FWHM could be adjusted; 
the observed profile followed by adding the individual $\tau$'s at every
pixel across the line.  Figure 6 shows fits to \ion{Ti}{1} $\lambda$4999,
where dashed lines outline the individual components, the solid line
the observed profile, and a series of crosses show the representation.
Table 2 gives the parameters found to fit 2 representative
\ion{T}{1} lines as well as \ion{Ba}{2} $\lambda$6496, \ion{Li}{1} $\lambda$6707,
and \ion{Rb}{1}
$\lambda\lambda$ 7800, 7947.

	 A source of uncertainty in these fits was the choice of
continuum level under the shell line.  Ideally that would be the underlying
photospheric line spectrum as represented by an artificially broadened
F-G supergiant (as in Fig.\ 5), but it was clear that an adequate match
could be achieved only by adding in a veiling continuum as well.  Until this
effect could be understood, and applied in a consistent wavelength-dependent
fashion, it was decided simply to interpolate the continuum level linearly
between two points just outside either edge of the shell line. The
equivalent widths of components 1 and 2 are particularly susceptible to
errors in the continuum level so defined.

	Nine unblended \ion{Ti}{1} lines having lower levels between 0.0
and 1.4 eV were synthesized in this way. For each of the 5
shell components, the EWs were fitted to a theoretical
pure-absorption curve of growth, and the values of the 
parameter $\xi_{0}$ (the Doppler width in velocity units), the
excitation temperature $T_{exc}$ and the
total \ion{Ti}{1} column density N(\ion{Ti}{1}) extracted.  They
are listed in Table 3.  These temperatures are estimated to be
uncertain by several hundred degrees. For most components $\xi_{0}$
ranges between about 2 and 5 km s$^{-1}$, intermediate between the
thermal velocities for Ti (1.1 km s$^{-1}$) and H (8 km s$^{-1}$) at
the T$_{exc}$'s of Table 3.  However the measured FWHMs of the individual
\ion{Ti}{1} shell lines scatter between about 8 and 19 km s$^{-1}$ (following
allowance for the instrumental FWHM of 6 km s$^{-1}$).  This shows
that there is another source of line broadening in these expanding
shells, possibly weak unresolved structure.

	Eight unblended \ion{Fe}{1} shell lines were analyzed in the same way. 
The scatter in the fit of the gaussian EWs to the curve of growth
was much larger than for \ion{Ti}{1}, possibly because of greater uncertainties
in defining the continuum level. Only the results for the $-78$ and $-61$
km s$^{-1}$ components are considered reliable.  The T$_{exc}$ for
\ion{Fe}{1} was very clearly lower than for \ion{Ti}{1}, between about 1350 and
1500 K.  The total column density if T$_{exc}$ = 1500 K is near
log N(\ion{Fe}{1}) = 17.4.  This difference between the T$_{exc}$'s of \ion{Ti}{1}
and \ion{Fe}{1} is puzzling, because their identical velocity structures
indicate that they originate in the same parcels of rising gas.  Possibly
non-LTE conditions in the shell are responsible.

	The evolution of the shell and wind absorptions during the brightness
minimum of 1996--2001 is shown in Figure 7, from SOFIN spectra.  One
might expect that the shell features would be stronger at minimum brightness,
but in fact at minimum in 1998--1999 it was the {\it wind} features that were
enhanced, while the shell components were strongest in 1996--97 and 2001.  This
is illustrated in Figure 8, which compares the 4660--4690 \AA\
region on the HIRES spectrogram of 1997 Aug.  12 (below) with that of 1998 Oct.
30 (above). The \ion{Ti}{1} shell lines (4667.58 and 4681.91 \AA) were strong on
the first date but had essentially disappeared 15 months later.  At that second
date, weak, narrow emission lines (arrowed) at approximately the stellar
velocity were detectable.

	An unusual feature of the shell spectrum is the prominence of the
lines of \ion{Rb}{1} at 7800 and 7947 \AA. Their EWs in the $-78$ km s$^{-1}$
component were 194 and 127 m\AA\ in 1997 August.  The first ionization
potential (IP) of Rb is 4.8 eV, so one would expect it to be ionized because
barium, with first IP = 5.2 eV and a comparable meteoritic abundance to
rubidium, is detectable only as \ion{Ba}{2} ($\lambda$6496, EW = 267 m\AA).
The non-detection of \ion{Cs}{1} $\lambda$8521 is understandable because of the
still lower first IP of cesium (3.9 eV) and a lower meteoritic abundance
(Rb/Cs = 19), but the strength of \ion{Rb}{1} remains unexplained.

\subsection{ The Emission Lines}

	Before 1997, the only obvious optical emission lines in V1057 Cyg
were the components of the P Cyg structure of H$\alpha$ and \ion{Ca}{2}. 
In that
year, when the star had begun its decline following the 1985--1994 plateau,
a number of emission lines of low excitation appeared in the centers of the
corresponding stellar absorptions \citep{pet98}.  The most
conspicuous were \ion{Fe}{1} [RMT 12] 8047 and 8074 \AA , followed by
\ion{Fe}{1} [60] 8514, \ion{Fe}{2} [40] 6516, \ion{Ca}{1} [1] 6572, and \ion{Fe}{1}
[12] 7912 \AA: Figure 9.  These emission peaks are at about the
stellar velocity, and are narrower than the photospheric absorption lines.
A  higher-resolution HIRES spectrogram of 1997 August 13 confirms the
asymmetry of the $\lambda\lambda$ 8047, 8074 lines that is apparent in
Figure 9: their shortward edges are clearly steeper than the longward.  Most of
these same low-excitation emission lines were observed long ago in the
G supergiant $\rho$ Cas by \citet{sar61}.

	This appearance of emission in the centers of many low-excitation
absorption lines is illustrated in Figure 10 which
compares the 6400 \AA\ region on a Lick coud\'{e} spectrogram of 1985 May 27
(resolution about 18 km s$^{-1}$) with the HIRES spectrogram of 1997 Aug. 13,
slightly smoothed.  In those intervening 12 years, the centers of the
\ion{Fe}{1} absorption lines $\lambda\lambda$ 6393, 6400 increased in
brightness with respect to the continuum, rising at peak to almost
the continuum level. (A second Lick spectrogram obtained on 1985 Sept. 25
showed no change had taken place during those 4 months.)

	We believe that this emission spectrum is produced in a warm layer
we call a `chromosphere' that is almost overwhelmed by the photospheric
continuum when the star is bright, except through its marginal appearance as
emission cores in lower-excitation stellar absorption lines.
However, that chromosphere must have become brighter by a factor of about
2 between 1985 and 1997.  The reason: between those dates V1057 Cyg faded by
only about 0.9 mag.\ in V, so if the 1997 emission cores in $\lambda\lambda$
6393, 6400 had been present at that same absolute brightness in 1985, they
would have filled those absorption lines up to about half the depth
actually observed.
	
	To determine whether there was any further change in the brightness
level of the chromosphere, the EWs of several emission cores were measured
in the SOFIN differential spectra (\S\ 2.5) of V1057 Cyg (minus $\beta$ Aqr
spun up to $v_{\rm eq}\, \sin i$ = 55 km s$^{-1}$ and not veiled) between 1996 and
2001.  The results are given in Table 4.  Over these years, the
emission line EWs remained constant within the errors of measurement;
i.e. as the star faded, the lines became weaker in the same proportion.
We conclude that there was no further change in the absolute intensity
of the chromosphere after the increase by a factor of approximately 2
sometime between 1985 and 1997.

	This brightening of the chromosphere may have been
related to the increase in the wind activity that began in 1984-85 (\S\ 2.2).

	As V1057 Cyg has declined in brightness, the fading of the continuum
has helped to reveal this chromospheric emission spectrum.  We believe that
it is
this spectrum that is responsible for the apparent `doubling' of
some absorption lines that has been suggested as evidence of orbital
motion in a Keplerian disk.  In an earlier paper on the spectrum of
FU Ori \citep{pet92} we argued for such an explanation of the
`doubling', a possibility in fact first mentioned by \citet{goo87}. The
fading of V1057 Cyg has now provided strong support for that
interpretation.

	These central reversals which have emerged as distinct emission lines
since 1995  are not very strong, e.g., \ion{Fe}{1} 
$\lambda$8047 has a peak intensity of only 5--8\% above the continuum level.
If the continuum were one magnitude brighter (or the chromosphere fainter),
the line would appear only as an emission core at the bottom of a broader
photospheric absorption, as in \ion{Li}{1} 6707 \AA\
(Fig.\ 9).  For a quantitative analysis of this emission spectrum, the
underlying photospheric contribution must first be subtracted. This is
described below (\S\ 2.5), as is a comparison with the low-temperature
(T$_{\rm exc}$ about 2700 K) emission spectrum of VY Tau at the time of a
flare-up, to which it bears a strong resemblance.

 	The central intensities of such chromospheric emissions would,
in the optically thick limit, rise to the Planck flux at that temperature.
Therefore---depending in individual cases upon {\it gf} value and lower
EP---the strengths of those lines with respect to the hotter
photospheric continuum would tend to increase toward longer wavelengths.
This would explain the greater prominence of line `doubling' in the red
than at shorter wavelengths that has been noted, for example, by HK.

	The strong \ion{Ca}{2} H \& K emission lines at 3933, 3968 \AA\ (discussed
in \S\ 4.5) and at the infrared triplet are probably produced in this
chromosphere, presumably the same that \citet{dan02a, dan02b} found
necessary to reproduce the H$\alpha$ profile of FU Ori.  The Balmer emission
lines that ought to be produced in the same region are concealed by the P Cyg
structure of the wind, except for the longward emission fringe at H$\alpha$. 
The shortward edges of the \ion{Ca}{2} emission lines are truncated by their own
P Cyg absorptions.  Clearly, the outflowing wind is located {\it above} this
chromosphere, demonstrated also by the presence of fluorescent \ion{Fe}{1}
4063 and 4132 \AA\ lines in V1057 Cyg: those \ion{Fe}{1} atoms ``see" the exciting
\ion{Ca}{2} $\lambda$3968 emission line, although it is hidden from us by the wind
component of H$\epsilon$.

	A very broad emission, at peak only about 0.12 above continuum level,
is present at 6297 \AA\ on the HIRES spectrum of 1997 August 13 (that region
falls between orders on other exposures).  It must be [\ion{O}{1}] $\lambda$ 6300.30
because the weaker [\ion{O}{1}] line at 6363 \AA\ is present on other HIRES and
SOFIN spectra of 1996--1998.  The central velocity of $\lambda$6300 is about
$-135$ km s$^{-1}$, its total width at continuum level about 180 km s$^{-1}$.
A similar broad, shortward-displaced emission line is also present at the
position of [\ion{Fe}{2}] $\lambda$7155.  They have nearly the same velocity as
the ``intermediate case between shell and wind" components mentioned in
\S\ 2.3.  No such features are found in FU Ori.

  	The reader should be aware that FUors are not unique in possessing
broad absorption lines with emission cores and CO absorption in the 2 $\mu$m
region:  a number of normal (i.e. not pre-main sequence) F- and G-type high
luminosity stars are known to have such spectra. Several examples were
mentioned by \citet{pet92} and more recently, the classical
case of $\rho$ Cas has been rediscussed by \citet{lob98}.

	To summarize, in addition to the photospheric spectrum, these
sets of spectral features were present in V1057 Cyg during this period at
different radial velocities:

\noindent
\hspace*{0.2in} wind at $-100$ to $-300$ km s$^{-1}$ (H$\alpha$, D$_{12}$ \ion{Na}{1},
 \ion{Ca}{2}, \ion{Mg}{1}, \ion{Fe}{1}I); \\
\hspace*{0.2in} shell at $-40$ to $-110$ km s$^{-1}$ (TiO and low excitation atomic
lines in absorption); \\
\hspace*{0.2in} CO molecules at about the stellar velocity (\S\ 3.5); and  \\
\hspace*{0.2in} low-excitation chromospheric emission at the stellar velocity.
  \\

\subsection{ The Differential Spectrum}

	As already pointed out, a number of low excitation lines clearly
went into emission above the continuum during the brightness minimum of
1996--2001.  It is natural that the same feature should be present in 
higher excitation lines if only as an emission core at the bottom
of the absorption line, but enough to cause those lines to appear
double and shallow, as is observed \citep{pet92, pet98}.
Such line emission can be revealed by subtracting
the underlying photospheric spectrum. As a template for the photospheric
spectrum of V1057 Cyg we use $\beta$ Aqr spun up to $v_{\rm eq}\, \sin i$ =
55 km s$^{-1}$ and veiled by a factor 0.3. Two fragments of this differential
spectrum are shown in Figure 11 for the average spectrum of 1998--2000. 
Note that the relative strength of the emission lines is not 
the same as in the absorption spectrum. 

	Both the ``true" emissions (that rise above the continuum)
and those revealed in such a differential spectrum fall along a common curve
of growth for T$_{\rm exc}$ = 3600 $\pm$ 300K, log n$_{e}$ = 7.5 $\pm$ 0.5.
This temperature is significantly lower than photospheric. In the M-type 
dwarf VY Tau, the same emission lines appeared very strong by contrast with
that low temperature continuum \citet{her90}.  There is a good correlation
between the equivalent widths of the low-temperature line emissions in
VY Tau and those in the differential spectrum of V1057 Cyg: Figure 12.

	The same method was used by \citet{wlt92}
to reveal the emission lines in the differential spectra of FUors, but 
the template they used for subtraction was the accretion disk model spectrum. 
The line depths in V1057 Cyg are now very different from both $\beta$ Aqr 
and that disk model (Fig.\ 13).

	Although the observed line intensities can be explained as a sum
of the photospheric and emission line spectra,
the observed {\it line profile} is not just a sum of two gaussians.
Weaker lines have a rather ``boxy" shape, with sharp edges, while stronger
lines (without shell components) have nearly normal rotational profiles
except for the emission cores.  This difference suggests some abnormality
in the structure of the lower atmosphere, deserving of attention at high
resolution $\ge$60,000 and S/N $\ge$ 300.

\section{ V1057 Cyg: Interpretation of the Light Curve }

	Conventional assumption is that a FUor outburst represents only
a temporary event and that the star will eventually return to its former
brightness. The slow fading of FU Ori over the past 60 years,
and the relatively rapid decline of V1057 Cyg since about 1971 might be
explained in this way.  But the spectrum of V1057 Cyg does not support
this expectation.  Before the 1970 outburst, the star possessed H$\alpha$
emission that, in order to have been detected at all in the first
low-resolution surveys, by \citet{har71} in the early 1950s and by
\citet{her58} in 1952--56, must have had an equivalent width of
approximately 25--40 \AA.  No such emission has appeared at H$\alpha$
during the decline to the 1999--2000 minimum: the emission fringe has
remained near EW(H$\alpha$) = 1--2 \AA\ (Table 1).  

	Another departure from expectation is the following.  Before the
outburst a number of \ion{Fe}{1} and \ion{Fe}{2} emission lines were reported
\citet{her58}, so they must have been fairly strong to have been detectable
on that 1957 low-dispersion photographic spectrogram.  Yet at the present
time no \ion{Fe}{2} emission lines are detectable in the blue-violet on
modern, far superior digital spectrograms, although the fluorescent
 \ion{Fe}{1} 4063, 4132 \AA\ lines are weakly present.

	Furthermore, as the star has declined one would have expected the
absorption spectrum to approach that of a TTS-like K- or M-type dwarf.
In 1998--1999 V1057 Cyg was only about 1.5 mag.\ (in B) above its pre-outburst
level (B = 14.9) so if the star was returning to its
original T Tau state, a late type photosphere and emission line spectrum
ought to have emerged.  But in spite of the drop in brightness, the
spectral type of V1057 Cyg remained the same as in the 1980's:  
in \S\ 2.1 it was shown that the star continues to resemble a
rapidly rotating G supergiant, unlike any other pre-main sequence star
of which we are aware. But the spectroscopic similarity is deceptive:
the M$_{V}$'s are quite different. If the surface brightness and
(V-R)$_{J}$ color of V1057 Cyg at the time of the 1985--1994 plateau
were the same as those of the standard G0 Ib $\beta$ Aqr, then correction
for A$_{V}$ = 2.35 mag.\ leads to M$_{V} = +0.3$ for a distance of 600 
pc.\footnote {This distance is based on the colors of stars in the general
vicinity of NGC 7000 \citep{her58, lau02}.}
This compares to M$_{V} = -3.5$ for $\beta$ Aqr.  That M$_{V}$ for V1057
Cyg would be produced by a single star of uniform surface brightness having
a radius of about 9 R$_{\odot}$.

	Given these considerations, and the fact that V1057 Cyg both
pre- and post-outburst was unusual in its possession of a very massive 
high-velocity wind (\S\ 4.5), we later speculate that pre-outburst FUors are
not normal TTS, but represent a special sub-species of that class.

	We suggest that the behavior of V1057 Cyg as it has faded, as well as
for the general spectroscopic properties of the classical FUors, may be
understood in terms of a stratified atmosphere atop a rapid rotator
with a strong quasi-permanent outflowing wind, a low effective $g$ being
responsible for the line-spectrum resemblance. The emission spectrum which
has appeared as the continuum of V1057 Cyg has faded is, as we have stressed,
that of a low-excitation chromosphere atop the stellar atmosphere (\S\ 2.4).

  	The decline of V1057 Cyg from its 1971 peak brightness to the plateau
level in
1985--94 can be represented by a continuous change in radius and surface
brightness of a rapid rotator, those quantities being derivable from the
procedure of \citet{bar76} and the above values of extinction
and distance.  Observed $V$ and $(V-R)_{J}$ for 1971--2001 were taken from
\citet{kop02}, for 1978--2001 from \citet[][and
private communication]{ibr96, ibr99}, and for 1971 from
\citet{men72} and \citet*{rie72}. The resulting values of R/R$_{\odot}$ are plotted in
Figure 14.  The symbols identifying the sources of the photometry are
explained in the caption.\footnote {\citet{kop84}
also used the Barnes, Evans, \& Parsons formulation to calculate
R/R$_{\odot}$, but from its dependence on $B-V$.  Our radii come instead from
the dependence on (V-R)$_{J}$, following the recommendation of Barnes et al.,
and the fact that an excess shortward of 4800 \AA\ was present between 1971
and 1975--76 \citep[][Fig.\ 8]{her77}, which would make $B-V$ suspect.  It is of
course unclear which color is more likely to be applicable to an unusual
object like V1057 Cyg.}   Radii and surface brightnesses depend upon whose
colors are used.  If the Ibrahimov data, Figure 14 shows how
R/R$_{\odot}$ fell from about 14 near maximum light to about 9 at the time of
the plateau.  The surface brightnesses declined from values appropriate a
late A-type main sequence star in 1971, to a mid-K type in 1995. 

	 We emphasize that such calculations {\it do not prove} that the source
is a spherical star, only that the brightness and color can be represented
by a circular surface of those dimensions and surface brightness.

  	In 1984--85 the wind flux began to increase (\S\ 2.2), and at about the
same time the decline in brightness halted.  Sometime between 1985 and 
1997---we speculate that it may have been early in that interval---the chromosphere
is known to have brightened with respect to the continuum.  Thus all three
phenomena may have been consequences of an upsurge of activity in the
underlying rapidly rotating star.

	The plateau episode ended when, between 1994 and 1995, the star
abruptly became fainter by 0.78 mag.\ in B, and redder by 0.18 mag.\ in $B-V$ 
(seasonal averages); see the small plot of B-V vs time at the bottom of Figure 1.
The ratio $\Delta B/\Delta(B-V) = 4.3$ is not far from the standard
interstellar reddening value (4.1).  Thereafter (\S\ 2.4) 
the continuum and chromosphere fluctuated in brightness together, so we ascribe
the 1994--95 fading to screening by a dust layer somewhere higher in the
atmosphere.  This is not a new idea: the
possibility of dust condensation in the outflowing wind of V1057 Cyg has
already been raised by \citet{kol97}, and it will be recalled that
\citet{ken91} interpreted a relatively brief dimming of the
FUor V1515 Cyg in 1980 as such an event. The formation of such a dust layer
was also envisioned by \citet{rao99} in the case of R CrB, also a
high-luminosity G star. 

	 However V1057 Cyg continued to fade, $\Delta B = 0.63$ mag.\ by 1999,
apparently without any further change in color.   Either more dust
dominated by large particles, or a continuation of the slow post-1971
decline, could be responsible.

	The foregoing is an attempt to pull together the photometric and
spectroscopic phenomena exhibited by V1057 Cyg since the 1970--71 outburst.
However, it is likely that the atmospheric structure is not radially
homogeneous, as this picture may seem to imply.  The day-to-day and secular
fluctuations
observed in the H$\alpha$ structure at both V1057 Cyg and FU Ori show that wind
ejection is spasmodic, possibly coming from localized areas on the rotating
star, rather as the fast solar wind emerges from `coronal holes' on the Sun.
It would then not be surprising if the wind fields above these stars
contained much structure. The HST images of V1057 Cyg (\S\ 3.1) show that
the distribution of dust near that star, whether formed in and ejected
by the star or local dust shaped by the stellar wind, is highly structured.

	The disappearance and reappearance of the shell spectrum of V1057
Cyg on a time scale of a few years cannot be due to pure radial
expansion and the consequent decline in column density proportional to
r$^{-2}$.  It may be caused by the movement of inhomogeneities in the
wind structure across the line of sight.  The broad shortward-shifted [\ion{O}{1}]
and [\ion{Fe}{2}] lines described in \S\ 2.4 could then arise in this expanding,
inhomogeneous envelope, their longward wings being occulted by the star.

	Unexplained is the veiling mentioned earlier that was used to account
for the general shallowing of the absorption spectrum.  It is conceivable that
the line shallowness is intrinsic, i.e., the result of integration over a
very non-uniform stellar hemisphere, or of line formation in a highly
non-spherical extended atmosphere, or of the contribution of a continuum
originating in the chromosphere.  If extrinsic, thermal emission by dust is
an unlikely explanation, because although the energy absorbed in the
hypothetical dust layer must reappear somewhere, dust would not survive at
temperatures greater than about 1500 K so that that re-emission would
be significant only at long wavelengths, not in the optical.

\subsection{ Direct Images of V1057 Cyg}

	If dust did form in the lower atmosphere of V1057 Cyg in 1994 and was
subsequently expelled with the wind, then in time
it might become detectable in scattered light.  Given an ejection velocity of 
200 km s$^{-1}$ and no subsequent deceleration, then the separation
of dust and star in the plane of the sky would increase at the rate of
0\farcs07 yr$^{-1}$, so that in the 5 years following 1994 that dust would
in projection appear about 0\farcs35 from the star.  

	Five WFPC2 images of V1057 Cyg are available in the HST archive.
They were obtained with HST on 1999 Oct. 18 as part of a `snapshot' program;
the filters were F606W (central wavelength 5957 \AA) and F814W (7940 \AA).
We are grateful to Karl Stapelfeldt, the Principal Investigator, for the
opportunity to study this material. We did no more than obtain the pipeline
processed images from the Archive and trim and clean up cosmic ray hits and
other defects.

        Those F606W images are shown in Figure 15.
There is much scattered light and spurious structure surrounding the
overexposed star image, so that nothing can be said whether structure
exists as near as 0\farcs3 from the star.  However, at least three features
at somewhat larger separations are present and appear to be
real, judging from inspection of similar images of ordinary stars on other
WFPC2 frames taken in the same series.  They are identified by letters.  C is
the brightest; it appears as a structureless blob protruding from the star
image to a distance of about 1$''$, while A and B are fainter, curved
arcs reminiscent of the larger loops at V1057 Cyg and other FUors
described by \citet{goo87}.   The arc D is more distant,
and is located at the base of a similar structure present on ground-based
images of V1057 Cyg obtained in the 1970s \citep{dun81}.
The reality of A, B and C is confirmed when the image of
a single star (from another frame) is subtracted from the shortest F606W
exposure: see the lower right panel of Figure 15.  The feature C is apparently
only a section of an extended nebulous bar.  There is no persuasive
correspondence of this structure very near the star with the molecular-line
or 1.3 mm continuum maps of \citet{mcm95}.

	 If C is a slab of warm dust
very near the star, consider the possibility that its thermal emission 
may contribute significantly to the integrated IR emission of
V1057 Cyg.  An estimate of its relative contribution can
be made as follows.   Assume that the true separation of star and slab is
as projected,  0\farcs8 (480 AU), and that the star radiates as a black
body of $T_{\rm eff} = 5300$ K and radius 9 R$_{\odot}$.  Then the equilibrium
temperature of a small silicate particle exposed to that radiation
field is obtained by balancing the energy absorbed by the amount
re-radiated.  Given the absorption cross-sections for `astronomical silicate'
\citep{dra85}, the temperature of a 0.1 $\mu$m silicate particle at that
position is found to be about 75 K, with a weak dependence on particle radius.
If the slab, an assemblage of such particles, radiates as a black surface of
dimensions 0\farcs33 $\times$ 0\farcs64 at 75 K,  then its flux would be
dominant over that of the star at wavelengths greater than about 14 $\mu$m.
On the other hand, if the slab preserved the optical properties of its
constituents, then the crossover wavelength would depend on particle radius,
being at about 19 $\mu$m for 0.1 $\mu$m particles, at 16.5 $\mu$m for
0.4 $\mu$m, and at 15.5 $\mu$m for 1.0 $\mu$m.  Obviously, thermal emission
from such nearby dust must contribute to the SED's of stars like V1057 Cyg,
although to lesser degree than these estimates if the cloud were optically
thin.

	Dust formed in the atmosphere of V1057 Cyg and then ejected 
could have reached the slab's
present position in as short as 12 years.  Nothing is known of the earlier
history of V1057 Cyg but if there have been previous outbursts, each with its
own dust-formation episode, such distant dust concentrations might be
explained.  Future high-resolution imaging will show whether these 
structures are moving with respect to V1057 Cyg.

	Unfortunately no comparable HST imagery is available at this time
for FU Ori.  Conventional ground-based CCD images obtained with the 2.3-m
telescope on Mauna Kea show extensive reflection nebulosity around that star,
with brightness increasing toward the star before merging at about 3$''$ with
the overexposed star image.  Coronagraphic images reproduced by \citet{nak95}
extend this in to about 2\farcs5.  As at V1057 Cyg, this
material and that detected at 2.3 $\mu$m very near FU Ori by \citet{mal98}
may contribute significantly to those SED's.

\section{ FU Ori, V1057 Cyg and FUors in General}
 
\subsection{  Comparison with the Composite Spectrum of an Accretion Disk}

	Accretion disk models were devised by \citet{ken88}
to explain the spectral energy distribution and the peculiar 
double-peak profiles of the photospheric lines in V1057 Cyg and FU Ori. 
In those models, the high-resolution spectra were synthesized by
assuming that at any given radius the disk radiates as a stellar 
atmosphere of the appropriate spectral type. The variation of
effective temperature with radius is given by the steady disk 
theory, and the variation of the rotational velocity with radius 
is assumed to be Keplerian. The integrated spectrum of the disk can 
then be represented as a composite of annuli of different temperature, 
surface brightness, and rotational velocity.
The models reproduce reasonably well the atomic lines in the optical 
region and the molecular bands in the infrared. In addition to FU Ori 
and V1057 Cyg, the spectrum of Z CMa was compared to the disk model by
\citet{wlt92}, revealing numerous emission lines in the
differential spectrum (i.e., observed minus synthetic) of Z CMa,
rather as is seen in V1057 Cyg near minimum brightness.

	However not all spectral lines can be reproduced by the
disk model.  Some lines in regions of the spectrum not examined by
\citeauthor{ken88} are in striking contradiction to the model
prediction. Because the spectrum of V1057 Cyg has changed due to the increasing
prominence of line emission as the star has faded toward minimum brightness,
in what follows we first demonstrate these discrepancies in the spectrum
of FU Ori because it has not changed significantly over the last two decades.  

	New synthetic disk spectra were calculated for FU Ori and V1057 Cyg
using the parameters of the disk models given by \citet{ken88}
and a set of our own template spectra (see Table 5)
obtained at the Nordic Optical Telescope with SOFIN.
As an example, the synthetic and the observed spectra of FU Ori are 
shown in Figure 16 for the spectral range 5260-5320 \AA . 
This synthetic spectrum looks identical to that shown in Figure 3 of
\citet{wlt92}.  The spectrum of FU Ori in 1998 was also very similar to
that displayed by \citeauthor{wlt92}. 

	In the accretion disk model for FU Ori, the relative contribution 
from different parts of the disk (i.e., from different spectral types)
to the total flux radiated by the disk depends on wavelength as is shown in
Table 6 [contribution].  At 5500 \AA\ the spectrum of the disk is mostly of
F-G type, while at 9000 \AA\ all spectral types contribute about equally. 
This means that at 9000 \AA\ one can find F-type spectral features
along with M-type, e.g., both high-excitation lines and TiO bands.

	If we consider relative line strengths, the composite spectrum
of the accretion disk looks much the same as the spectrum of a normal 
G supergiant because the same atomic lines are changing smoothly from 
late F through G to early K types. The difference can be found only
in early type F, where the lines of high-excitation species
appear strongly,  and in type M where molecular bands, 
mostly of TiO, appear very strong. Since both F and M spectral types
contribute to the accretion disk model of FU Ori, we examine
the observed spectrum in order to determine how these critical features
behave.

	The most suitable spectral region for such an analysis is around
8900 \AA. It contains two \ion{Ca}{2} lines having lower EP
of 7.05 eV (8912.06 and 8927.35 \AA ) which are very strong only in type F, 
a line of V I at 8919.80 \AA\ having EP = 1.2 eV which increases in strength
from type G to M, and a strong TiO bandhead at 8860 \AA, 
characteristic of type M. 

 	The telluric spectrum was extracted from a spectrum of the O7e
fast rotator $\xi$ Per, and with it weak terrestrial lines were
removed from the FUor spectra. In order to reduce the noise,  
all the spectra of FU Ori taken in 1997--2000 were averaged. 
For V1057 Cyg, to avoid a possible contribution from the shell, only the
1998--2000 spectra were averaged because in those years the shortward-shifted
TiO features were absent.

	Comparison of the synthetic spectrum of the accretion disk and
the observed spectrum of FU Ori is shown in Figure 17.  Some of the
template spectra are also displayed in the Figure to show the origin
of the main features in the synthetic spectrum. As expected, the synthetic
spectrum of the accretion disk shows all the F- and M-type features.
In the observed spectrum of FU Ori the blend at 8860 \AA\ resembles that
in the synthetic spectrum, except that the TiO head is not obviously present
probably on account of the overlap by Paschen line at 8862 \AA .  The absence
of TiO is better shown in V1057 Cyg (Fig.\ 21, below).   More 
obvious is the difference between the observed and synthetic spectra in the
width of the \ion{Ca}{2} and \ion{V}{1} lines.  That region is expanded in
Figure 18.  In the synthetic spectrum these lines have very different 
widths because they originate from the innermost and from the outermost 
regions of the disk, respectively rotating at very different Keplerian
velocities, but in the observed spectrum the lines have about the same widths.

	Another example is shown in Figure 19: the
high-excitation line of \ion{N}{1} $\lambda$7442.30, EP = 10.3 eV, is strong in
the F-type spectra, is present and highly doubled in the synthetic spectrum
of FU Ori but is absent in the observed one.  Figure 20 shows that
the ScO/TiO feature at 6036 \AA\ is absent in the observed spectrum of FU Ori,
while it is quite obvious in the synthetic.

	The 8900 \AA\ spectral region in V1057 Cyg also reveals a difference
between observed and model spectra (Fig.\ 21). Since the
rotational velocity is lower than in FU Ori, the TiO head is well resolved in 
the model spectrum. The TiO feature is clearly absent in the average 
spectrum of V1057 Cyg, or in the noisier 1998--2000 individual spectra.

	Here we have considered only those spectral features detectable at
the resolution of the SOFIN material. Much more could be done with spectra
of higher resolution. But at the moment we conclude that comparison
of the observed and synthetic spectra of FU Ori in the near infrared shows
that some critical spectral features do not have the structure predicted
by the multi-temperature disk model, although some elaboration of that
model might be able to account for the mismatches. A rapidly rotating
single star will have a latitude-dependent spectrum, whose appearance in
integrated light will be a function of aspect angle.  It remains to
be seen if such an object might produce a FUor-like spectrum.

\subsection{ Rotational Line Widths}

	The value of $v_{\rm eq}\, \sin i$ and the nature of the photospheric line
broadening in FUors have been a subject of much discussion.  \citet{wlt90,wlt92} 
measured absorption line widths on V1057 Cyg spectrograms obtained in 1986 and 1988
when the star was brighter, and found them to depend on wavelength, being larger in the
blue and smaller in the red. This they interpreted as demonstration of
differential rotation in a Keplerian accretion disk, as predicted by
the HK disk model.

	The shortward-displaced shell spectrum had been present at some level
on all our spectra of V1057 Cyg, but is most prominent at the shorter
wavelengths where most of the low-level lines of the neutral metals are
located.  It was very strong in 1996--97, so $v_{\rm eq}\, \sin i$ was measured
only on the 1998--2000 SOFIN spectra and as an additional precaution,
to avoid any marginal shell contribution, only the longward wings of
the photospheric lines were fitted.  Thirty selected lines were compared to
the template spectrum ($\beta$ Aqr, G0 Ib) spun up to a set of discrete values
of $v_{\rm eq}\, \sin i$.  Since the line depth in V1057 Cyg at that time was smaller
than in the template, a veiling contribution (which does not affect the
line width) was applied to rescale the line
depth.   

	The result was that in V1057 Cyg the average $v_{\rm eq}\, \sin i =
 55$ km s$^{-1}$, with a scatter of individual lines mostly between 50 and 60
km s$^{-1}$. No systematic trend with wavelength was present in the range
5000-9000 \AA, nor was there any trend with EP in the range 1--8 eV.  The
minimal level of veiling is about 0.3. It is larger for individual lines
filled in with emission, especially those of lower EP (see below).

	This value of $v_{\rm eq}\, \sin i$ (55 km s$^{-1}$) is larger than those
published by \citet{wlt90} (35--45 km s$^{-1}$). However, they measured
line half-widths at half depth, which in a rotationally broadened profile
is about 0.8 $v_{\rm eq}\, \sin i$ (the precise value of the factor depending
slightly upon the limb darkening assumed). We have also measured the
half-widths of the red wings at half depth: the average value is
44 $\pm$ 4 km s$^{-1}$, in
agreement with Welty et al., but again we found no dependence on wavelength or
EP.   It is possible, of course, that the spectrum of V1057 Cyg changed in
this respect during the intervening decade.

	The same procedure described above (fit of longward wings to the
spun-up template) was carried out for FU Ori.  The best fit was with
$v_{\rm eq}\, \sin i$ = 70 km s$^{-1}$, and again no relationship between
$v_{\rm eq}\, \sin i$ and EP or wavelength was found.  It would not seem likely
that the spectrum of FU Ori has changed significantly since the
Welty et al.\ observations.

\subsection{ CO Lines}

	On 1999 Oct. 24 both V1057 Cyg and FU Ori were observed in the
2.3 $\mu$m region by K. Hinkle with the Phoenix infrared spectrometer
\citep{hin98} at the KPNO\footnote {KPNO is operated by the Association
of Universities for Research in Astronomy, Inc. under cooperative agreement
with the National Science Foundation.}  2.1 m
telescope.   The region covered was 4320 to 4340 cm$^{-1}$ (2.315 to 2.304
$\mu$m) in the 2-0 CO band.  These spectrograms had been obtained at
the request of Lee Hartmann, and we are grateful to him for access to
the data and to Ken Hinkle for supplying us with the reduced spectra,
as well as those of several early-type stars observed on the same
occasion.

	This region contains many terrestrial lines, mainly of CH$_{4}$.
These are not removed completely by standard star  division because
of airmass mismatches.  Instead, the value of optical thickness
(required to reproduce the line depth as exp(-$\tau$)) was
calculated at every pixel across these features in the standards, which
was then scaled to cancel those features in the FUor spectra.
These spectra of FU Ori and V1057 Cyg are shown in the upper rows of Figure 22. 
The dominant features are the broad R-branch lines of the CO 2-0 band (R(19)
through R(27): lower EP 0.09 to 0.18 eV).  Their equivalent widths are
comparable to those of early to middle-M type giants and supergiants in the
atlas of \citet{wal96},
from which the FTS spectrum of $\lambda$ Dra (type M0 III) in the
bottom section of Figure 22 has been extracted.  The same spectrum spun up
to $v_{\rm eq}\, \sin i = 60$ km s$^{-1}$ is shown just above.
In $\lambda$ Dra the lines of the returning R-branch (R(75) through R(81):
lower EP 1.33 to 1.55 eV) are strong, but they are not apparent in
either FUor.  Although they would be masked to some
degree by the greater line widths, the lack of convincing and consistent
asymmetries in the FUor lines due to these blends suggest a lower
rotational temperature.

	The CO lines have different shapes in the two FUors: in FU Ori they
are broad and essentially symmetric, while in V1057 Cyg they are asymmetric,
with a narrow core and an extended wing toward negative velocities.  To
reduce the effects of instrumental noise and that introduced by atmospheric
line removal, the line profiles shown in Figure 23 were
created by averaging the 4-5 least noisy CO lines for each star.  These were
fitted by a procedure similar to that described in \S\ 2.3. The instrumental
profile,  represented by that of a narrow atmospheric CH$_{4}$ line of
FWHM = 8 km s$^{-1}$, could be spun up to adjustable $v_{\rm eq}\, \sin i$,
central velocity, and central depth.  The parameters of the fits
are given in Table 7.  Although the off-center structure in FU Ori may
be real, a single component serves to represent the overall profile reasonably
well.  In the case of V1057 Cyg, the observed profile is certainly
{\it not} a narrower version of the stellar absorption lines. Formally, it
can be fitted to two overlapping, rotationally broadened copies of the
atmospheric line.  Because of the complex structure, comparison of these
$v_{\rm eq}\, \sin i$'s with the optical value cannot be very meaningful. 

	In neither star is there any evidence of shell structure at large
negative velocities, perhaps understandably because the shell spectrum in
the optical was very weak at the time of the CO observation.  The CO velocity
of FU Ori agrees closely with the conventional velocity of that star
(+28 km s$^{-1}$), in agreement with \citet{mou78}, who found ``from several CO and
metal lines" a value of +28.2 $\pm$ 2.5 km s$^{-1}$. However, the velocities of both
CO components of V1057 Cyg (Table 7) are displaced from the optical $-16$ km s$^{-1}$,
nor do they agree with the Mould et al. value of $-13.4 \pm 3.5$ km s$^{-1}$.

	In the case of FU Ori, the value of $v_{\rm eq}\, \sin i$ from CO
(48 km s$^{-1}$) is clearly smaller than that we obtain from the
optical region (70 km s$^{-1}$), in the same sense as earlier work
by the CfA group.  \citet{har87a} measured total widths of
cross-correlation peaks rather than $v_{\rm eq}\, \sin i$'s, so the two results
cannot be compared directly, except to say that they
found the CO lines in FU Ori to be narrower than the optical by a factor of
about 0.75,  as compared to our 0.69.   In V1057 Cyg also, the stronger
CO component (45 km s$^{-1}$) is  narrower than the optical total
$v_{\rm eq}\, \sin i$ of 55 km s$^{-1}$, but this may not be significant 
because the CO profiles are peculiar. The only published CfA result for CO
in V1057 Cyg is an estimate that $v_{\rm eq}\, \sin i$ is about 20 km s$^{-1}$
\citep{har87b}.

\subsection{ Search for Rotational Modulation in Line Profiles}

	In T Tauri stars (TTS), a period of axial rotation can be derived
from rotational
modulations of brightness and line profiles caused by surface 
inhomogeneities (cool or hot spots) and nonaxisymmetric structure 
of the wind \citep[see, e.g., ][]{pet01}.
Unlike an ordinary star, an accretion disk has no single rotational
period but a range of Keplerian orbital periods. In the accretion disk
models for FU Ori and V1057 Cyg \citep{ken88}, 
the Keplerian periods range from 3--4 days for the inner disk where the
F-type spectrum is formed, to 30--40 days for the outer regions contributing
to the M-type spectrum. Thus an observational test would be to search for
rotational modulation in the wind, emission and photospheric line profiles. 
The most prominent wind feature is the H$\alpha$ P Cyg line.
The emission lines of metals are present only in the spectrum of V1057 Cyg 
and are too weak to be used in such an analysis because the individual 
spectra of V1057 Cyg near minimum brightness are rather noisy. 
The photospheric lines, although also being weak, can however be analyzed
by cross-correlation, which compensates for the poor signal-to-noise ratio
of the individual spectra.

\subsubsection{Wind Profiles}

	Figures 24 and 25 show overplotted all the H$\alpha$ profiles from
the SOFIN data of 1995--2001 for V1057 Cyg and FU Ori, the latter in
two parts to emphasize the major changes in the longward emission
component.  The left panel is a superposition of all the profiles of
FU Ori in which the emission peak intensity was greater than 1.20 (in
continuum units), and the right panel all those with peak less than 1.10.
As already mentioned, there is no obvious correlation between these
emission peak intensities and the extension of the P Cyg absorption
component to negative velocities.

	The time scale of the variability is different in the two objects. 
In V1057 Cyg, the night-to-night variability is relatively small, but the 
average profile is quite different in different years (see also Fig.\ 7). 
In both objects, most variable is the central emission peak and the 
portion of the profile between about $-100$ and $-400$ km s$^{-1}$.

	The same variations are seen in the P Cyg absorptions of the
\ion{Na}{1} D$_{1,2}$ lines, which correlate well with H$\alpha$, although the
\ion{Na}{1} lines have a somewhat smaller range in wind velocity.  The
absorption
at the infrared \ion{Ca}{2} lines also vary together with H$\alpha$ and
 \ion{Na}{1}
but the velocity amplitude is much smaller.  H$\alpha$ was chosen for
analysis because it showed the largest range in wind velocity.
 
	In FU Ori, the line profiles change considerably on a time scale
of a day \citep[as was noted most recently by][who give
earlier references]{dan02a}.  The following variability patterns are seen in the 
H$\alpha$ profile on SOFIN spectra:  (1) There is no correlation between the
changes of the absorption and emission components (we return to this
matter in the following section).  (2) Three
sections of the absorption component vary independently of one another:
 (a) a ``slow wind" at RV = $-50$ to $-110$ km s$^{-1}$,
 (b) a ``fast wind" at $-110$ to $-270$ km s$^{-1}$,
 (c) an even faster wind at $-270$ to $-400$ km s$^{-1}$.
(3) The fast wind portion shows quasi-periodic
variations in EW, but (4) no periodicity is found in the emission component
of H$\alpha$.

  	The dense time coverage of the SOFIN spectroscopy makes feasible a
search for periodicity, so the EW of the P Cyg absorption
between $-110$ and $-270$ km s$^{-1}$ was used as a parameter of the
fast wind of FU Ori.  Those EWs are listed in Table 8. 
The phase dispersion minimization method \citep{ste78} was used to
search for periodicity in these data. When the velocities of only three
seasons,  1997 to 1999, are examined (20 nights), the periodogram (Fig.\ 26,
upper panel) shows a group of significant peaks at 13--18 days, with the most
probable value being 14.847 days.  Fisher's method of randomization
\citep{nem85} gives a false alarm probability (FAP)
of $<$0.01 for that period.  There are no other significant periods
between 2 and 100 days.  The phase diagram (Fig.\ 26, lower panel)
shows that that period is defined largely by the 1997 and 1998 data.  The data
of 1999 also fit, but span a shorter phase interval.  The periodicity
is not present in the data of 2000 (7 nights)  although the fast wind EW
varied over almost the same range.

\citet{err03} have recently published H$\alpha$ profiles of
FU Ori obtained over 5 consecutive nights in January 1999.  The
variations in the P Cyg absorption (measured by them as velocity width at an
intermediate depth) indicated that if periodicity was present, the period
could be about 6--8 days, shorter by a factor 2 than found from our more
extended set of EW measurements.  Considering that our observations in 
2000 show no periodicity, it may be that the 14.8 day cycle had already
become undetectable in 1999.  Our own data of October 1999 (4 nights)
are too limited to pronounce upon the presence of a 6--8 day cycle in that
year.  We conclude that the variations of the wind absorption in FU Ori
are quasi-periodic: in 1997--98 the period was 14.8 days but that periodicity
had disappeared by 2000.

	\citeauthor*{err03} suggested that these cyclic variations in the wind
of FU Ori are caused by the interaction of the stellar magnetic field with 
the disk outflow, assuming that the magnetic axis of the star is inclined to
the disk rotational axis.  However, in that model of an inclined magnetic
rotator one would expect rather stable periodicity on a very long time scale.
Instead, the {\it quasi-periodic} character of the variations is more
like that observed in TTS, where surface inhomogeneities (spots) may appear
and disappear due to changes in the star's magnetic field structure. 

	We suggest that the longer period of 14.8 days suggested by the
H$\alpha$ variations (because that line showed the largest velocity
amplitude)  is the rotational period of FU Ori.  In case the wind is governed
by the star's magnetic field (as
is believed to be the case in TTS), a relatively stable axial asymmetry of
the field structure might cause the observed rotational modulation over at
least two years (1997-98), or 50 rotational cycles.  The observed
$v_{\rm eq}\, \sin i = 70$ km s$^{-1}$ then leads to a value of R sin $i$ = 
20.4 R$_{\odot}$.

	In the accretion disk model for FU Ori (\citet{ken88})
a period of 14.8 days corresponds to a Keplerian orbit of radius
20 R$_{\odot}$, where the K-type spectrum is supposed to be formed, but 
it is not clear why a global disk wind should be modulated with any 
single period.

	If the intrinsic optical colors of FU Ori are those of a normal
G0 Ib, then the observed values for the 2000 season
($V = 9.58$, $V-R_{J} = +1.185$)  correspond to
$E(V-R)_{J} = 0.575$, so if the reddening is normal, the average $A_{V}$ 
is 1.38 mag., leading to $M_{V} = -0.1$ for the assumed distance of 450 pc.
The visual flux in solar units is

\begin{equation}
	 \frac{F_{V}}{F_{V}(\odot)} = 2.512^{-M_{V}+M_{V}(\odot)} 
\end{equation}

If only a fraction $x$ of the surface is occupied by regions having the
G0 Ib brightness, then the radius of such a spherical star is

\begin{equation}
       \frac{R}{R_{\odot}} =
 ( \frac {F_{V}/F_{V}(\odot)}{xb} ) ^{1/2}   
\end{equation}

\noindent
where $b$ (= 0.7) is the ratio of surface brightnesses at V calculated by the
procedure of \citet{bar76}.  If $x$ = 1, the result is that
R = 11.3 R$_{\odot}$.

	Consider whether this last result can be reconciled with
R sin $i$ = 20.4 R$_{\odot}$. If the star's surface brightness is not
uniform, then any value of the coverage factor $x$ $\le$ 0.32 could
bring the two into agreement, the inclination then following from sin $i$ =
(20.4/11.3)$x^{1/2}$.

	Another possibility: if the A$_{V}$ inferred from the color excess
of FU Ori were ignored and instead taken to be 2.66 mag., regarded as the sum of
conventional interstellar reddening in the foreground and a circumstellar
component of unspecified reddening properties, then M$_{V}$ would become
-1.3 and R = 20.4 R$_{\odot}$, in agreement with the axial rotation result
for sin $i$ = 1.  If the entire A$_{V}$ was circumstellar, then  A$_{V}$/E(B-V)
= 5.4, compared to the normal interstellar value of 3.1.  A/E ratios greater
than about 4.0 are not unprecedented, having been found in dusty \ion{H}{2} regions
and in dense molecular clouds \citep{car89, chi98},
there usually being ascribed to the presence of larger particles.  That may
not be inconceivable in the case of FU Ori because the local reddening of
such a dusty object need not obey the normal interstellar extinction law.
But there must surely be a normal interstellar contribution to A$_{V}$, in
which case the A/E ratio of the circumstellar component would become even
larger, as it also would if sin $i$ $<$ 1.0 for FU Ori.

	So either hypothesis could reconcile the two results.  We see no
strong observational reason to favor one hypothesis over the other at this
time, except that the dust explanation does require circumstellar extinction
of unusual properties.  The periodic variation in wind strength observed in
FU Ori, with its
implication of unevenly distributed active areas, favors the blotchy surface
hypothesis.  If precise photometry showed that FU Ori varies cyclically with
that same period, it would strengthen that proposition.  Dust formation low
in the atmosphere, leading to a larger A$_{V}$ and an abnormal A/E and
subsequent ejection, might explain dust structure very near the star.  But at
this time the question remains open.

\subsubsection{Photospheric Profiles}

	As remarked earlier, the double-peaked profiles of (some) photospheric
lines have been used as one of the arguments in support of the accretion-disk
model of FUors (HK).  So far, no spectral time series have been available to
determine how stable is this peculiar line structure, which we attribute
(\S\ 2.4) to the presence of emission cores in the low-excitation lines.
In our spectral series of V1057 Cyg and FU Ori (SOFIN data of 1995--2001)
this line structure is found to be variable on a time scale of several days.
In the following we use the cross-correlation method to
give information about line profiles averaged over a certain spectral
region, usually one spectral order.  The spectrum of $\beta$ Aqr (G0 Ib)
is used as a template.

	As an example, Figures 27 and 28  
show night-to-night variability of the cross-correlation functions (CCF's)
during one set of observations of V1057 Cyg and FU Ori. The CCF profile varies
synchronously in different spectral regions: the line width at half-depth
remains about the same, while the central part varies with either shortward or
longward peak being stronger, sometimes becoming single and quite symmetric.
Of course the CCF profile depends on the particular mix of low- and 
high-excitation lines in the sample, and this may be why different spectral
orders show somewhat different CCF shapes,  but the variations are similar
in different orders.  The spectral intervals selected for cross-correlation
were chosen to avoid lines with shell components.

	As a descriptor of line position we use the ``center of gravity" of
the CCF, namely its weighted mean radial velocity.  This is the
velocity of the star as if it had been measured from the ``center of gravity" of
a photospheric line profile, although variations of this quantity
do not necessarily mean that the whole star moves around in radial
velocity.

        In the case of V1057 Cyg two spectral orders (6320--6440 and 6575--6640
\AA) were combined; the resulting precision of RV is about $\pm$3
km s$^{-1}$.  The spectra of FU Ori are of better quality, so  6
spectral orders (from 5560 to 7520 \AA) were used, the resulting
precision being $\pm$1 km s$^{-1}$.  Table 9 contains the measured RV's
for both stars.

	 Since the individual CCFs are quite
noisy, especially for V1057 Cyg, three groups of CCFs showing
similar RVs were selected: shortward-shifted, centered, and longward-shifted.
The three CCFs shown in Figure 29 are averages for these three
groups, overplotted to illustrate typical variations of the CCF profile.
For V1057 Cyg each CCF is an average of 8 spectra, for FU Ori each is an
average of four.  Obviously, the main source of the variability is the
deformation of the central part of the line profile: the ratio of
shortward-to-longward peak intensity is variable, while the width of the line
remains about the same.  This latter fact excludes the possibility of a
double-line binary, where the line width must be narrower when the two
components have the same radial velocity.

	For FU Ori, the periodogram shown in the upper panel of Figure 30 
was calculated using the data of the 1997, 1998 and 1999 seasons
(20 nights). The most probable period is 3.542 days, with
FAP $<$ 0.01.
When phased with this period, the lower panel of Figure 30 shows the cyclic
variation of RV in these three years.  The semi-amplitude of the sinusoidal
variations of RV is 1.8 km s$^{-1}$.  The scatter of points around the
sinusoidal curve is less than $\pm$1 km s$^{-1}$, which must be entirely
due to the errors in RV. When all the data of 1995--2000 are used (29 nights),
the period is still present but the data of 2000 are not well fit to the
sinusoidal curve. It is concluded that the period of 3.542 days in the
photospheric lines of FU Ori was stable during at least three years.

	For V1057 Cyg, using all the data of 1995--2001 (42 nights),
the periodogram reveals a group of peaks around 4.4--4.5 days, 
with the most probable period being 4.43 days.  However, the semi-amplitude 
of the RV variations (3 km s$^{-1}$) is comparable to the errors in RV
(3 km s$^{-1}$), which makes it unlikely that the periodicity is real,
so we draw no conclusions from this result.

	The nature of such variations in the photospheric line structure
is not clear.  In the case of a single star, variations like those shown in 
Figure 29  can be caused by an asymmetric polar starspot.
However, a dark spot must also modulate the apparent stellar brightness.  No
such variation has been reported for FU Ori. Since the photospheric line
doubling is due to the presence of emission cores, variations in the
doubling may be caused by movement of those cores. It is also
possible that the line emission originates from a volume of gas 
distributed non-axisymmetrically around the star. Such ``emission spots"
could also produce a rotational modulation of the photospheric line profile.
On the other hand, if the wind cycle of 14.8 days is the rotational period,
it is hard to explain why the period of the photospheric variations is 4
times shorter.

	In case of the accretion disk model \citep{ken88},
a period of 3.542 days corresponds to Keplerian rotation
at a distance of 7.7 R$_{\odot}$ (if the mass of the central star
is 0.5 M$_{\odot}$).  This is the innermost part of the disk,
where the F-type spectrum is supposed to be formed. If variability of the
photospheric lines is caused by some kind of brightness asymmetry in the disk,
the stability of such an asymmetry over three years (300 orbital periods)
would be difficult to understand in a differentially rotating disk.

	Apart from the deformation of the line profiles, small shifts of the
CCF can be noticed in V1057 Cyg (Fig. 29): the change in the
shortward-to-longward peak intensity is accompanied by  shifts of the entire
CCF profile by a few km s$^{-1}$. If these shifts are real, and not an artifact
of cross-correlating noisy spectra, it could indicate the presence
of a low mass secondary near the star. Series of better quality spectra are
needed to check whether this effect is really present.

	\citet{unr98} have inferred the presence of
hotspots on the rotating TTS DF Tau by an analysis of such cyclic deformations
of absorption line profiles.  Those hotspots, which they suggest are
accretion shocks at mass-infall points,  are clearly hotter than any of those
on the classical FUors, where there is no sign of the \ion{He}{1} lines at
5875 or 7065 \AA\ in emission.

\subsection{Infall}

	Our new spectroscopic material refutes a concern
that we raised in 1992, namely that there was then no direct
spectroscopic evidence of infall in FUors, such as ``reversed P Cyg"
structure. Disk theory yields an
accretion rate onto the FUor central star of about 10$^{-4}$ M/M$_{\odot}$
yr$^{-1}$ \citep{har96}, as compared to $\approx$10$^{-7}$
M/M$_{\odot}$ yr$^{-1}$ for classical T Tauri stars (CTTS.  One would think
that the movement
of such massive amounts of material on to the star would surely be detectable.

	The disk hypothesis has been extensively elaborated since HK and
a number of possible explanations have been offered for the lack of evidence
for infall: (a) much of the accreted material may be
ejected from the disk surface as wind, and so never reaches the star; or
(b) the accreting mass may accumulate in the disk;
or (c) the central source may be so faint that absorption lines would not
be detectable against its continuum; or (d) the disk may be so thick at its
inner edge that any activity nearer the star is concealed (in fact
\citeauthor{kle96}  [\citeyear{kle96}]
predict that at an accretion rate of 10$^{-4}$ M/M$_{\odot}$ yr$^{-1}$
"the entire stellar surface is engulfed by opaque disk gas"); or (e) an 
optically thin infall region may be hidden not because of disk thickness
but because these FUors are observed at unfavorable aspect angles. 

	However, as a result of the detailed time coverage at SOFIN,
evidence of what appears to be sporadic infall onto the continuum source
can now be seen in the H$\alpha$ profiles of both FU Ori and V1057 Cyg.
On two (of 29) SOFIN spectra a weak absorption component appeared at about
RV = +90 km s$^{-1}$ in the longward wing of H$\alpha$.  Figure 31 shows
three of these spectra, demonstrating that the feature was present on
JD 2450796.68, but not at comparable strength on the
day before or two days later.  The pattern was the same at H$\beta$, and the
absorption was marginally detectable at the \ion{Na}{1} D$_{12}$ lines, but
not at the \ion{O}{1} $\lambda$ 7773 blend which in CTTS is sensitive to
accretion.  A similar absorption had appeared at H$\alpha$ at about the same
velocity on
JD 2450807.69, but had been absent on the day before.  Clearly, these infall
features in FU Ori vary rapidly in strength, on a time scale of one day or
less.  The longward wing of the absorption in H$\alpha$  extends to about
+150 km s$^{-1}$, which is near the free fall velocity at the surface of a
star of 1 solar mass and a radius of 20 R$_{\odot}$. A similar event has
been seen in a HIRES observation of FU Ori in early 2003. 

	It is uncertain if the complete suppression of the longward
component of H$\alpha$ as seen in the right-hand panel of Figure 25
can be ascribed to very heavy accretion of this kind. 

	Thus infalling material does appear sporadically
in the line of sight to both FU Ori and V1057 Cyg, as is observed in
many CTTS  \citep{edw94}.  It is there thought to be due to
magnetically-channeled, free-falling disk material.  A signature of such
magnetospheric accretion is the presence of emission lines of \ion{He}{1}
and \ion{He}{2} that are believed to originate in the hot spots where
infalling material impacts the star \citep{ber01}.  As already noted,
no such emission lines appear in any of our spectra of the classical FUors. 

	It has been suggested that in CTTS a consequence of such infall
may be the ejection of Herbig-Haro-like jets, and it is interesting that
the spacing of structure in some H-H outflows does correspond to estimates
of the time spacing of repetitive FUor events in TTS.  A signature of
such shocked gas is the presence of lines of [\ion{O}{1}], [\ion{S}{2}] and
sometimes [\ion{N}{2}] \citep{cab90}.  All that we have
found in the integrated spectra of the FUors that we have observed are the
very weak, broad [\ion{O}{1}] and [\ion{Fe}{2}] emissions in V1057 Cyg
(\S\ 2.4).
They have large negative velocities and so must be formed in the outflowing
wind.  The [\ion{S}{2}] or [\ion{N}{2}] lines are not detected, although they
would be expected to be much weaker.   However, no H-H-like jets
are seen in the direct images of the FUors we have observed, although
a search at higher angular resolution with the proper filters would be
worthwhile.

\subsection{ The \ion{Ca}{2} H and K lines in FUors}

	An unusual feature of the pre-outburst spectrum of V1057 Cyg
was that although the K line ($\lambda$ 3933) of \ion{Ca}{2} was strong in emission,
the H line ($\lambda$3968) was absent.  The obvious explanation of this oddity,
as was realized long ago, is that $\lambda$3968 is quenched by the P Cyg
absorption component of H$\epsilon$ $\lambda$3970.  The important conclusions
are (a) that 12 years {\it before} the 1970 flare-up, V1057 Cyg was subject to
a strong mass outflow: i.e. the high-velocity wind did not turn on at the
time of the outburst; and (b) that wind was seen against the spectrum of the
pre-outburst star.
 
	The same anomaly was also present 28 years {\it after} the flare-up:
a HIRES spectrogram obtained 1998 October 30, when the star was about 1 mag.\
(in B) above minimum brightness, is shown in Fig.\ 32.  It
demonstrates not only that the H$\epsilon$ wind is indeed responsible for
the K:H anomaly, but implies that such a wind may be a quasi-permanent
characteristic of the object.
	
	This same wind suppression of \ion{Ca}{2} $\lambda$3968 has been
observed
in FU Ori since the time of the first adequate spectroscopy (1948). It is
seen not only in the other classical FUor that we have observed (V1515 Cyg) but
also in Z CMa which has been called a FUor, as well as in V1331 Cyg to which
attention was called long ago by \citet{wel71} for that very reason.  Only BBW76
\citep{rei90, rei97, rei02} does not conform: narrow emission is present
in both \ion{Ca}{2} lines.  Figure 32 shows HIRES spectra of the 3900--3980 \AA\
region
for all 6 stars.  The presence of strong \ion{Ca}{2} emission in all the
FUors, both near or below maximum light, shows that a permanent
chromosphere is characteristic of the group.

	This K:H anomaly is not a common feature of TTS spectra.
Most TTS outflows occur at modest negative velocities such that the
absorption component at H$\alpha$ falls within the broad underlying
emission line: see the atlases of H$\alpha$ profiles by \citet{fer95}
and by \citet{rei96}.  FUor outflows are dramatically different, in
that the mass involved is much larger and extends to much greater
negative velocities, hence the characteristic P Cyg profile and the
suppression of H$\epsilon$.  If this K:H anomaly is indeed a FUor
signature, can it be that there are unrecognized FUors among the host
of ordinary TTS?  There are 65 stars with \ion{Ca}{2} emission in
the atlas of TTS spectra by \citet{val93} and of these
only 3 (AS 353A, LkH$\alpha$ 321, V1331 Cyg) have the
\ion{Ca}{2} $\lambda$3968 line suppressed.
The spectra of 3 additional stars apparently showing the same effect
have been published by \citet{per01}.  Detection
of such stars by slitless spectroscopy would be an efficient means of 
searching for new FUor candidates.
	
\subsection{  Does the FUor Phenomenon Occur in Every T Tauri Star?} 

	Before the 1970 outburst, V1057 Cyg was only one of some fifty faint
H$\alpha$-emission stars scattered over the NGC 7000-IC 5070 region.
Since it seemed to be just another TTS---on the slender evidence
of that single pre-outburst low-resolution slit spectrogram---it was suggested in
\citet{her77} and then again in \citet{her89} that such outbursts might be a
characteristic of TTS in general.  If so, from the number of such outbursts
that had been detected (3 at that time) and a guess as to how many TTS exist
within an observable distance around the Sun, it was estimated that ``the mean
time between successive FU Ori-like outbursts in an individual T Tau star" is
about 10$^{4}$ years.  That estimate has since been refined by others
but the basic concept has survived. 

	As appealing as that idea is, and the way it can be worked into a
larger picture of pre-main sequence evolution \citep{har96}, it was
a pure conjecture.  Consider these points:

       First, if every TTS is a potential and presumably recurrent FUor,
then FUors would be expected to appear in clusters or associations rich in TTS.
None of the classical FUors occur in the Orion Nebula, around
$\rho$ Oph, or in other regions containing large numbers of TTS.  V1057 Cyg
lies in an isolated dark cloud containing only one other very faint H$\alpha$
emitter,\footnote {It is the star arrowed in Figure 1a of \citet{dun81}. }
no other TTS have been found in the elongated streamer northwest of IC 5146
where V1735 Cyg is located, while V1515 Cyg is one of several H$\alpha$
emitters scattered over an extended obscured region.  BBW76 is alone in
a small dark cloud, with no known TTS in the vicinity.  There are exceptions,
however: there are a number of TTS near the small dark cloud B35 in which FU
Ori lies, and the heavily obscured L1551 IRS 5 is located in a loose
grouping of TTS at the southern edge of the Taurus clouds.  Another
possible exception may be the candidate FUor CB34V which, according to
\citet{alv95}, appears to lie on the near side of a heavily obscured
aggregate of young stars.

	Second, it was pointed out above (\S\ 4.6) that at minimum light,
some 12 years before the outburst, V1057 Cyg exhibited the same strong
outflowing wind that it has shown ever since the flare-up.  The same P Cyg
structure at H$\alpha$ is found in other FUors observed near maximum light.
Such vigorous mass ejection is not a property of ordinary TTS, which is the
reason it has become to be regarded as an indicator of FUor group membership.

	Third, others have noted that FUors tend to have large values of
$v_{\rm eq}\, \sin i$.  Even higher than the $v_{eq}$ sin $i$'s found here
for V1057 Cyg (55 km s$^{-1}$) and FU Ori (70 km s$^{-1}$) is the
$~$ 145 $\pm$ 20 km s$^{-1}$ reported by \citet{alv97}
for the candidate FUor CB34V.  FUors clearly have significantly
higher rotational velocities than most TTS, although a minority of TTS
are also rapid rotators:  four of some 60 TTS listed by \citet{har89}
and by \citet{bou86} have $v_{\rm eq}\, \sin i > 50$ km s$^{-1}$.   We are
encouraged to believe that rapid rotation is an essential characteristic of
the FUor phenomenon.

	There is another property that, at first, would
seem to distinguish FUors from most conventional TTS.  \citet{goo87}
noted that all the FUors recognized at that time were located at the
edge of either a complete loop or ring of reflection nebulosity; in some
cases only a section of the loop is seen as a curved arc terminating at
the star.  In addition to the well-known outer arcs at V1057 Cyg, the
HST images in Figure 15 show two curved filaments (A and B) that could
be part of such a recently ejected loop.  Interestingly, V1331 Cyg and
Z CMa (both of which show the K:H anomaly), as well as BBW76 are also
attached to such nebulae.  But so are seemingly unrelated stars like SU
Aur and AB Aur ($v_{\rm eq}\, \sin i$ about 65 km s$^{-1}$ and 80 km s$^{-1}$,
respectively), showing that the presence of such a nebula is not exclusively
a FUor signature, but may rather be the consequence of a very large $v_{\rm eq}$.

	These reasons encourage the speculation that FUor events may
not be a property of ordinary TTS, but may be confined to a special
rapidly rotating sub-species among them. But if the FUor phenomenon is in
fact favored in a certain kind of pre-main sequence object, it raises
the question: what could be special about stars that so often occur
in isolation, in contrast to those formed in groups or clusters? 

\section{ Summary and Conclusions}

	In \citet{pet92}, following a study of a high-resolution
spectrogram of FU Ori obtained in 1987, we asserted that that 
spectrum could be reproduced by an emission-line shell (which in the
present paper we call a `chromosphere') overlain by a rising, cooler absorbing
layer (now the `shell') atop a G-type supergiant absorption spectrum whose
lines were broadened by macroturbulence + rotation.  It was also pointed out
that some of the spectroscopic properties of FUors that have been argued as
proof of the accretion disk picture (line doubling,
the presence of CO bands in the near infrared, the possession of an IR excess)
are found in a number of much older high-luminosity stars where there is no
reason whatsoever to think that an accretion disk is present.

	The new SOFIN and HIRES spectroscopy was intended to reexamine
those assertions.  With respect to the three issues critical to the accretion
disk hypothesis that were raised in \S\ 1, our conclusions are: 

	(a) The emission cores responsible for the `doubling' of low-excitation
absorption lines that have become more apparent as V1057 Cyg has faded
clearly are produced in a low-temperature chromosphere.  We believe that
it is they that caused those absorption lines to appear double when the star
was brighter.  Presumably such chromospheres are present in the other
classical FUors as
well, judging from the ubiquity of \ion{Ca}{2} $\lambda$3933 emission.

	(b) We have been unable to confirm, with superior spectroscopic
material, the dependence of $v_{\rm eq}\, \sin i$ upon either wavelength or
lower EP that has been urged as evidence for the accretion disk model.
However we do agree that the CO lines in the 2 $\mu$m region of FU Ori are
narrower than are optical lines shortward of 0.9 $\mu$m; the situation for
V1057 Cyg is uncertain.   Whether this is unarguable
support for the disk hypothesis is not clear: it would be interesting to
determine if the same effect appears in the CO lines of normal G
supergiants.

	(c) Synthetic disk spectra that resemble FUors at optical wavelengths
fail to match certain critical lines in the near infrared, indicating that
syntheses such as those described in \S\ 4.1 can no longer be regarded as
firm support for the disk hypothesis, although some elaboration of that
hypothesis might reduce the discrepancies.

	The point is that we find that some of the {\it observational}
evidence that has been offered in support of the disk hypothesis can
either be explained in another way or cannot be confirmed. Although
details are somewhat different, the proposal of \citet{pet92}
and of \citet{her89} is strengthened: namely, that the observable
properties of the classic FUors are better explained by a single star
rotating near the limit of stability and losing mass via a
powerful wind, as was originally proposed by \citet{lar80}.

	On slender evidence, V1057 Cyg is assumed to have been a
classical TTS before the 1970 outburst.  Although by 1999 it had
declined to about 1.5 mag.\ (in B) above minimum, the spectrum still
(by 2002) has not begun to resemble that of a conventional CTTS.  In time it
may do so, and some of the issues raised here may be resolved.  Meanwhile, it
is important that the star be monitored spectroscopically on a regular
basis.

\acknowledgements

	We are much indebted to the late V. Shevchenko and to M. Ibrahimov for
providing us with their FUor photometry, to Keith Budge, Lee Hartmann, Ken
Hinkle and Karl Stapelfeldt for material which we have found very useful
in the course of this investigation, to Ted Simon for assistance with
the CO data, to Ilya Ilyin for assistance with the SOFIN observations, and
to Bo Reipurth for helpful comments.
The work of G.H. on FUors has been partially supported by the
U.S. National Science Foundation under Grants AST 97-30934 and 02-04021.
The work of P.P. and R.D.
was supported by the EC Human Capital and Mobility Network project ``Late-type
Stars: activity, magnetism, turbulence," and of P.P. by the CAUP grant
``Financiamento Plurianual."

\clearpage
\singlespace
\begin{figure}
\epsscale{1.0}
\plotone{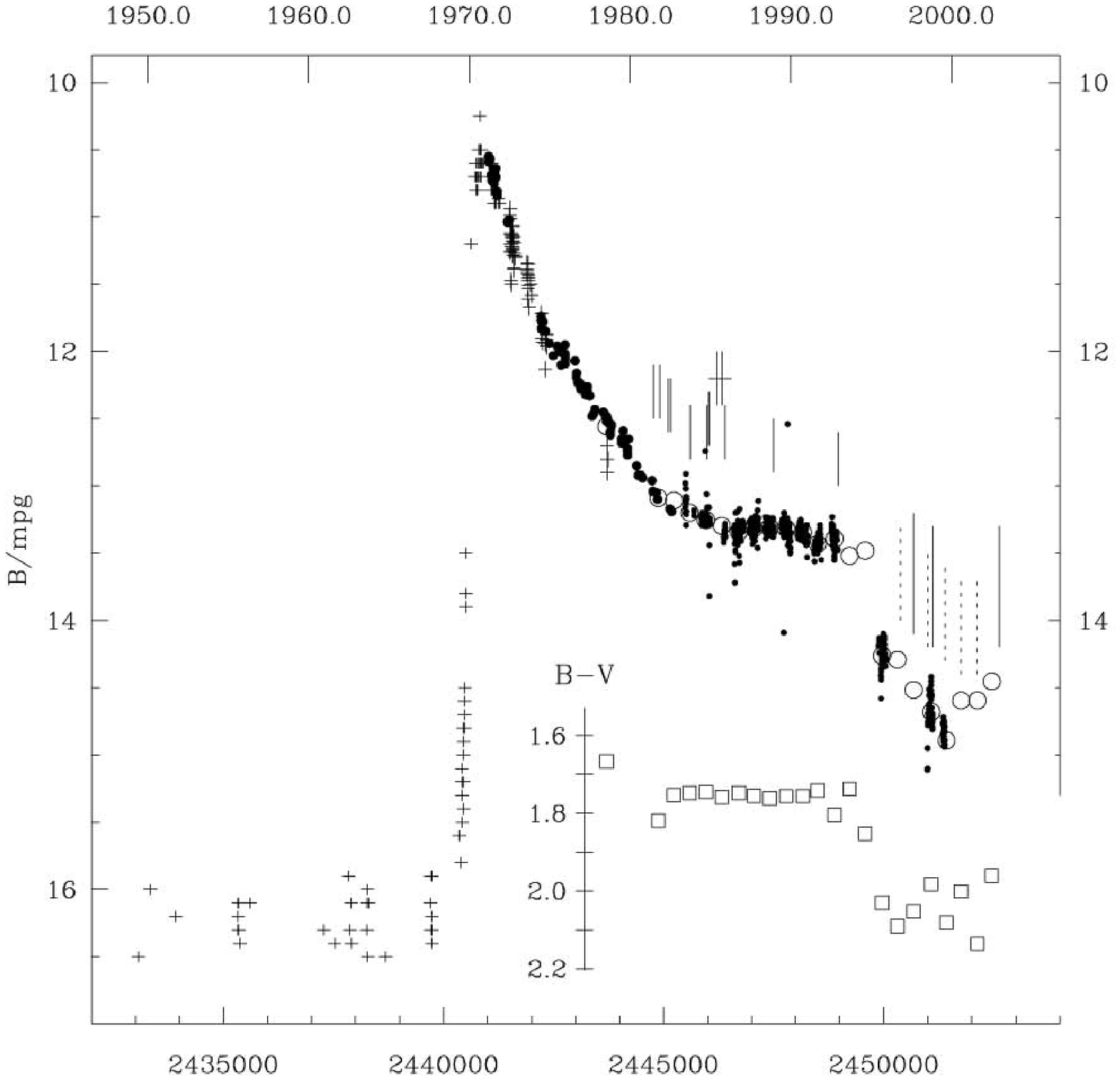}
\caption{ The B/pg light curve of V1057 Cyg.  Crosses represent photographic
observations, solid points are photoelectric or CCD magnitudes, open circles
are seasonal averages (from \citealt[][and private
communication]{ibr96,ibr99}). The vertical lines mark the dates of
high-resolution spectroscopic observations:  short lines by the CfA group or others at KPNO,
long dashed lines by Petrov with NOT/SOFIN, long solid lines by Herbig
with HIRES.   The two short lines with crossbars represent the dates of the
two Lick spectrograms of 1985.  The lower section shows, as small squares,
the seasonal average $B-V$'s, again from Ibrahimov.  The increase in reddening
in 1994-95 coincided with an abrupt drop in brightness.
  \label{fig1} }
\end{figure}
%\clearpage

\begin{figure}
\epsscale{1.0}
\plotone{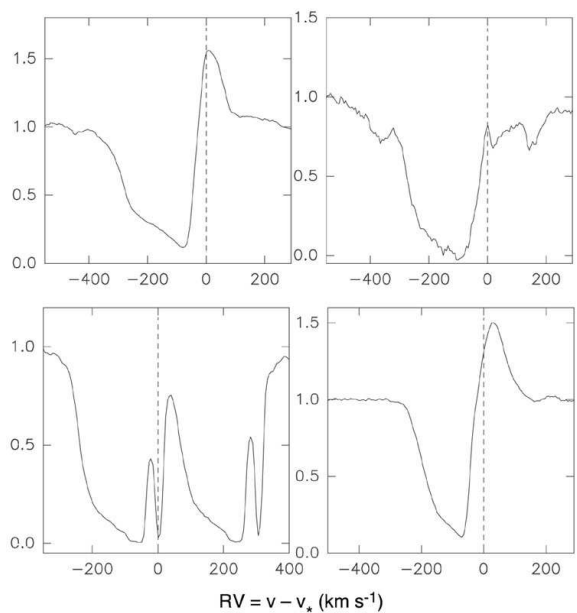}
\caption{P Cyg structure of the H$\alpha$ (upper left), H$\beta$ (upper right),
\ion{Na}{1} D$_{1,2}$ (lower left) and \ion{Ca}{2} $\lambda$8542 (lower right) lines
in V1057 Cyg from the SOFIN spectra of 1997 August.
   \label{fig2} }
\end{figure}
%\clearpage

\begin{figure}
\epsscale{1.0}
\plotone{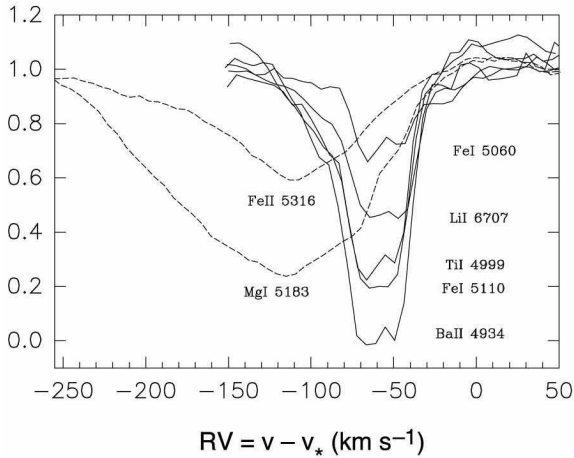}
\caption{V1057 Cyg shell lines in 1997, from SOFIN spectra of 13 km s$^{-1}$
resolution.  The spectrum of $\beta$ Aqr, spun up to 55 km s$^{-1}$ and
veiled by a factor 0.3, has been subtracted.  The velocity scale is in
the stellar rest frame.  Low excitation lines (EP $<$1 eV) are shown by
solid lines, higher excitation lines (2-3 eV) are dashed.
The strong lines of higher excitation potential, as \ion{Mg}{1} $\lambda$5183
and \ion{Fe}{2} $\lambda$5316,  have as well shortward-shifted components at large
velocities, about $-120$ km s$^{-1}$. 
   \label{fig3} }
\end{figure}
%\clearpage

\begin{figure}
\epsscale{1.0}
\plotone{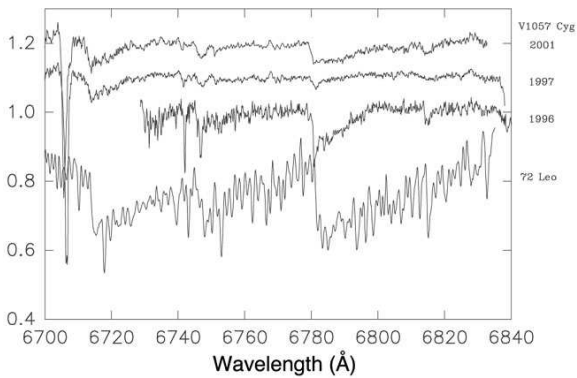}
\caption {The shell TiO bands in the spectrum of V1057 Cyg in 1996, 1997, and
2001.  The photospheric spectrum has been subtracted. The strong absorption
near 6706 is the \ion{Li}{1} shell component. The lowermost curve is the 
spectrum of 72 Leo (M3 IIb).
  \label{fig4} }
\end{figure}

%\clearpage

\begin{figure}
\epsscale{1.0}
\plotone{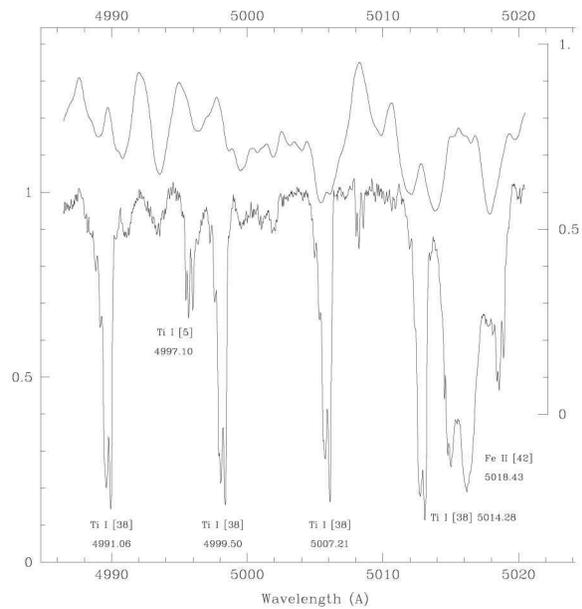}
\caption{The 4990--5020 \AA\ region in V1057 Cyg, from the HIRES spectrogram
of 1997 August 12 (scale on the left), showing the strong shell lines
of \ion{Ti}{1}.  Square brackets enclose their RMT multiplet numbers.  The upper
panel (scale on the right) is the same region in HD 190113 (type G5 Ib)
broadened by $v_{\rm eq}\, \sin i = 55$ km s$^{-1}$.
   \label{fig5} }
\end{figure}
%\clearpage

\begin{figure}
\plotone{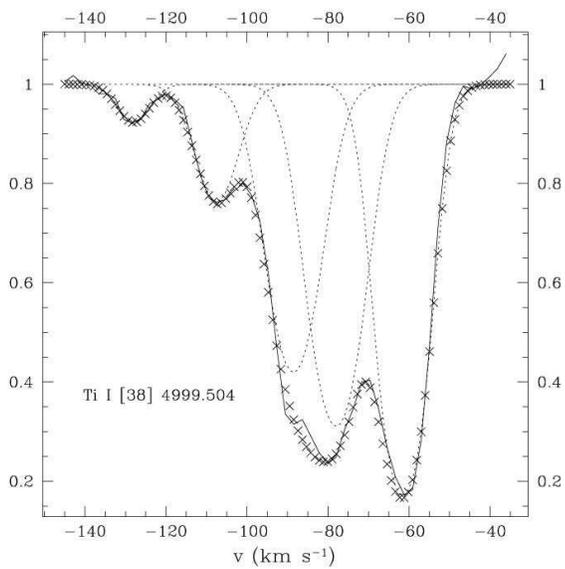}
\caption{Example of the resolution of the shell components of \ion{Ti}{1}
$\lambda$4999.504 by the procedure of \S\ 2.3.  The dotted lines outline the
gaussians fitted to each of the 5 components, the crosses their sum, the
solid line the observed profile.
  \label{fig6} }
\end{figure}

%\clearpage

\begin{figure}
\epsscale{1.0}
\plotone{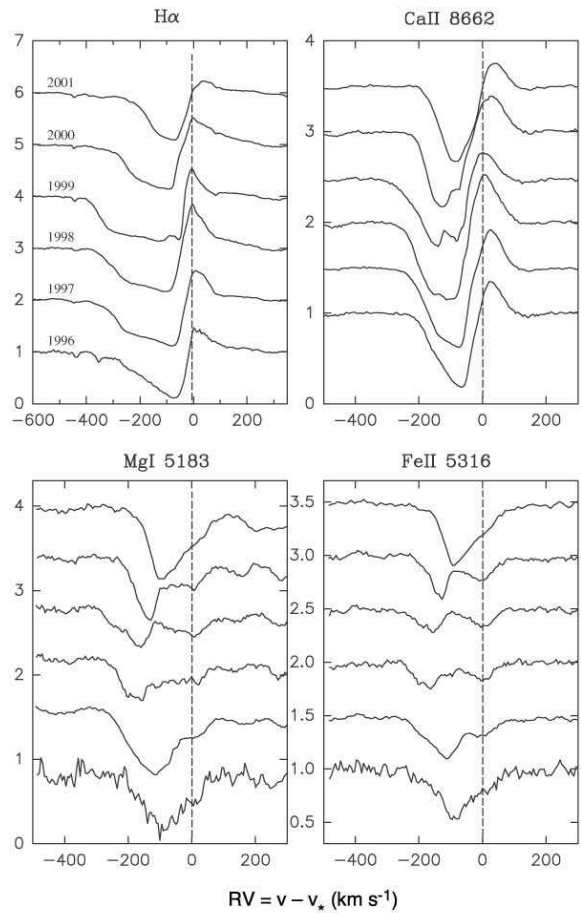}
\caption{Star and shell line profiles during the 1996--2001 brightness
minimum of V1057 Cyg.
   \label{fig7} }
\end{figure}
%\clearpage

\begin{figure}
\epsscale{1.0}
\plotone{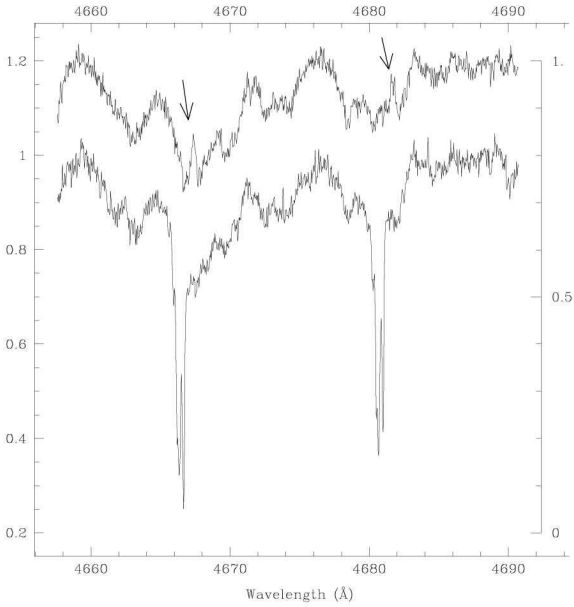}
\caption{The 4660--4690 \AA\ region in V1057 Cyg from HIRES spectrograms
of 1997 August 12 (below, scale on the left) and 1998 October 30 (above, scale on
the right).  Between those dates
the shell spectrum had essentially disappeared, illustrated here by the
4667 and 4681 \AA\ shell lines of \ion{Ti}{1}.  In these and many others,
narrow emission components (here arrowed) at approximately the stellar
velocity became apparent when the complex shell absorption lines
vanished.
  \label{fig8} }
\end{figure}
%\clearpage

\begin{figure}
\epsscale{1.0}
\plotone{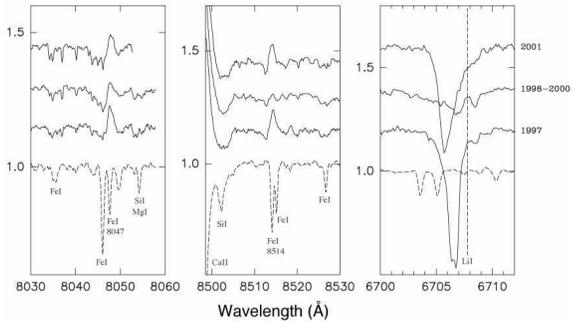}
\caption{Emission lines of \ion{Fe}{1} $\lambda$8047 and \ion{Fe}{1} $\lambda$8514 in
V1057 Cyg at three epochs (solid). The template spectrum of $\beta$ Aqr is
at the bottom (dotted).  The vertical line marks the position of the \ion{Li}{1} 
$\lambda$ 6707 line in the V1057 Cyg stellar rest frame. 
   \label{fig9} }
\end{figure}
%\clearpage

\begin{figure}
\epsscale{1.0}
\plotone{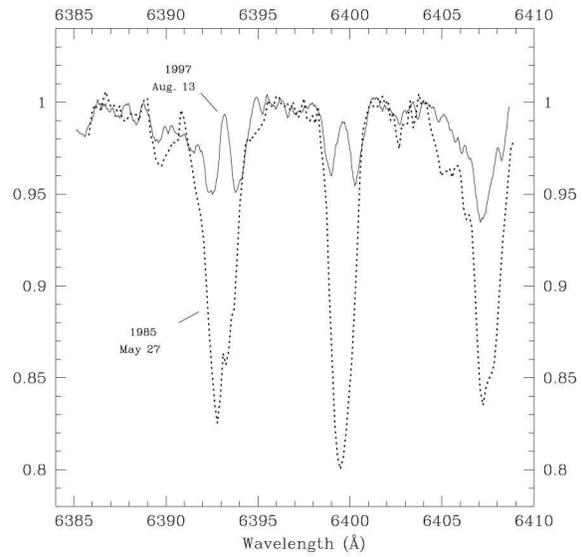}
\caption{ The 6385--6408 \AA\ region in V1057 Cyg, from a Lick CCD spectrogram
of 1985 May 27 (dotted) and from a HIRES spectrogram of 1997 August 13 (solid
line);  the latter has been smoothed by a 3-pixel box.  In the intervening 12
years the centers of the 6393, 6400 \AA\ \ion{Fe}{1} absorptions have become
emission lines, with peak intensities nearly at continuum level. 
  \label{fig10} }
\end{figure}
%\clearpage

\begin{figure}
\epsscale{1.0}
\plotone{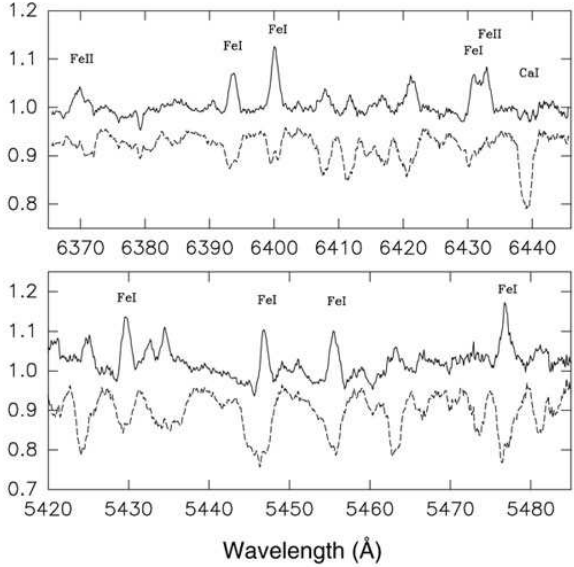}
\caption{ Two sections of the differential spectrum (solid line) obtained as 
a difference betwen that of V1057 Cyg (dashed line) and the template spectrum 
of $\beta$ Aqr, spun up to 55 km s$^{-1}$ and veiled by 0.3.
   \label{fig11} }
\end{figure}
%\clearpage

\begin{figure}
\epsscale{1.0}
\plotone{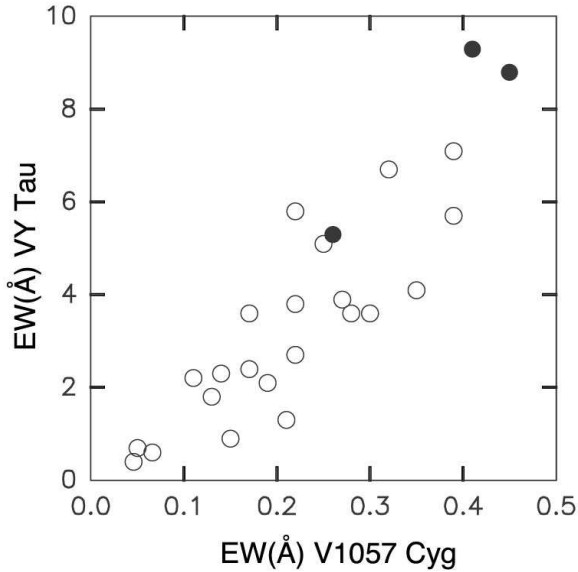}
\caption{ Correlation between equivalent widths of emission lines in
V1057 Cyg and VY Tau. Filled circles: the ``true" emission lines
seen above the continuum level in V1057 Cyg. Open circles: the emission 
lines revealed in the differential spectrum of V1057 Cyg.
   \label{fig12} }
\end{figure}

\newpage
 
\begin{figure}
\epsscale{1.0}
\plotone{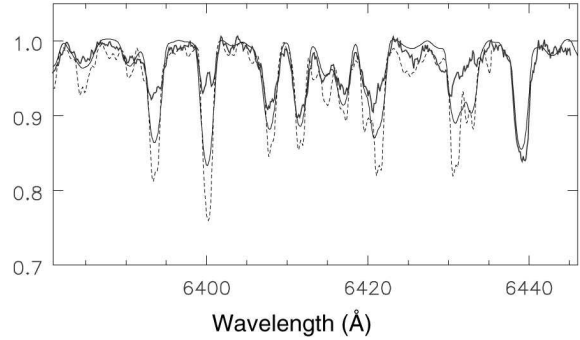}
\caption{ Photospheric lines in the average spectrum of V1057 Cyg in
1998--2000 (solid line). Shown for comparison are the spectrum of 
$\beta$ Aqr, spun up to $v_{\rm eq}\, \sin i$ = 55 km s$^{-1}$ and veiled by 0.3
(thin line), and the synthetic spectrum of the accretion disk (dashed line), 
calculated according to \citet{ken88}.
   \label{fig13} }
\end{figure}

%\clearpage

\begin{figure}
\epsscale{1.0}
\plotone{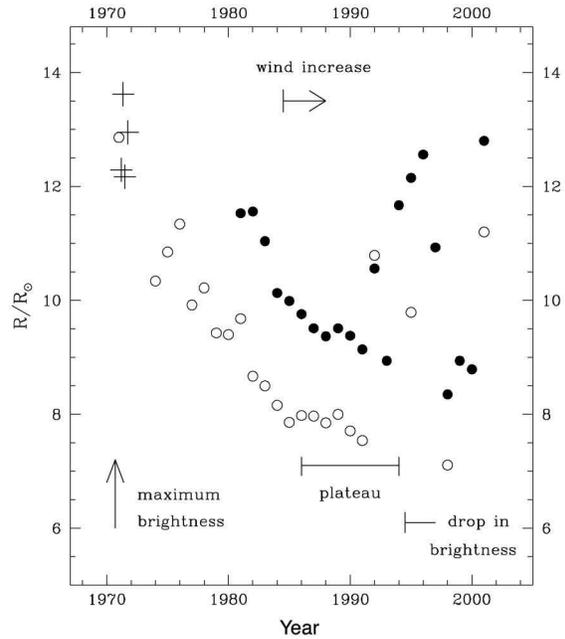}
\caption{The variation of radius (in solar units) of V1057 Cyg from
maximum light until 2001, calculated from V, (V-R)$_{J}$ observations
of \citet{kop02}: open circles; \citet[][priv.\ comm.]{ibr96,ibr99}: 
filled circles; and
\citet{men72} and \citet{rie72}: crosses.
	 \label{fig14} }
\end{figure}

\clearpage

\begin{figure}
\epsscale{1.0}
\plotone{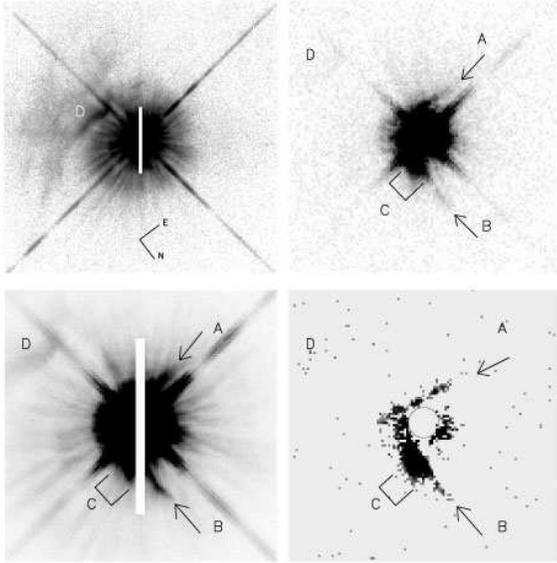}
\caption{Hubble Space Telescope images
of V1057 Cyg, obtained with WFPC2 on 1999 October 18 and filter F606W
(central wavelength 5957 \AA).  Upper left: an average of two 180 s exposures
on a logarithmic intensity scale. The white bar conceals the detector
bleeding of the star image; it is 3\farcs5 long.  The field size is 14\farcs6
square.  Upper right: an area of 5\farcs5 on a side, centered on the star, from
the shortest (14 s) exposure.  Lower left: the same image and scale, but
intensity scaled to emphasize the faint outer nebulosity (D).  Features very
near the star that are believed to be real are identified by letters
A, B, C.  Lower right:  the previous frame after subtraction of an image of
a single star from another WFPC2 exposure taken in the same series.  The
white circle outlines the saturated central region.  The diffraction spikes
and much of the scattered light structure has thereby been cancelled, showing
the underlying nebulosity.
 \label{fig15} }
\end{figure}

%\newpage

\begin{figure}
%\epsscale{1.0}
\plotone{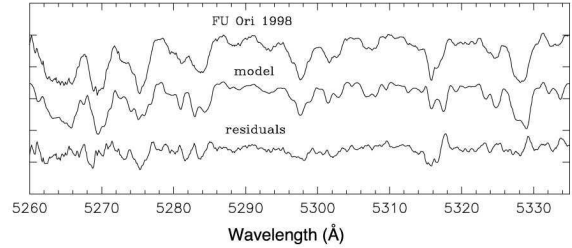}
\caption{Comparison between the observed spectrum of FU Ori and the 
synthetic spectrum of the accretion disk model.
	\label{fig16} }
\end{figure}

%\clearpage

\begin{figure}
%\epsscale{1.0}
\plotone{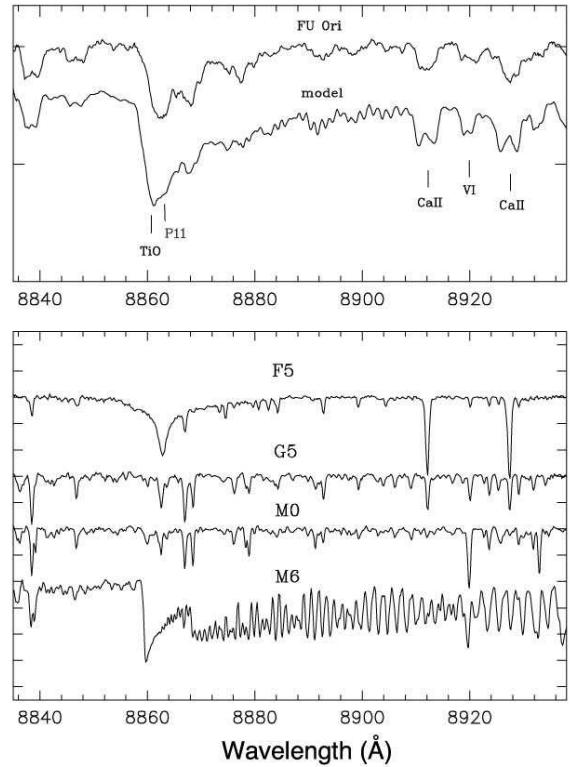}
\caption{ Upper panel: Comparison between the spectrum of FU Ori and the 
synthetic spectrum of the accretion disk model for the region of the
TiO band and the IR \ion{Ca}{2} lines. Lower panel:  examples of the template 
spectra used in calculations of the accretion disk spectrum. 
	\label{fig17} }
\end{figure}
%\clearpage

\begin{figure}
%\epsscale{1.0}
\plotone{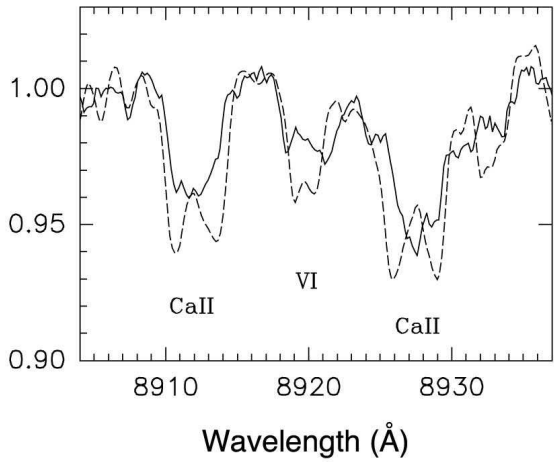}
\caption{ As Fig.\ 17 but for the region of the IR \ion{Ca}{2} lines.
 Solid line: observed spectrum of FU Ori. Dotted line: synthetic spectrum
 of the accretion disk. Note the differences in the line widths and profiles.
	\label{fig18} }
\end{figure}

%\clearpage

\begin{figure}
%\epsscale{0.75}
\plotone{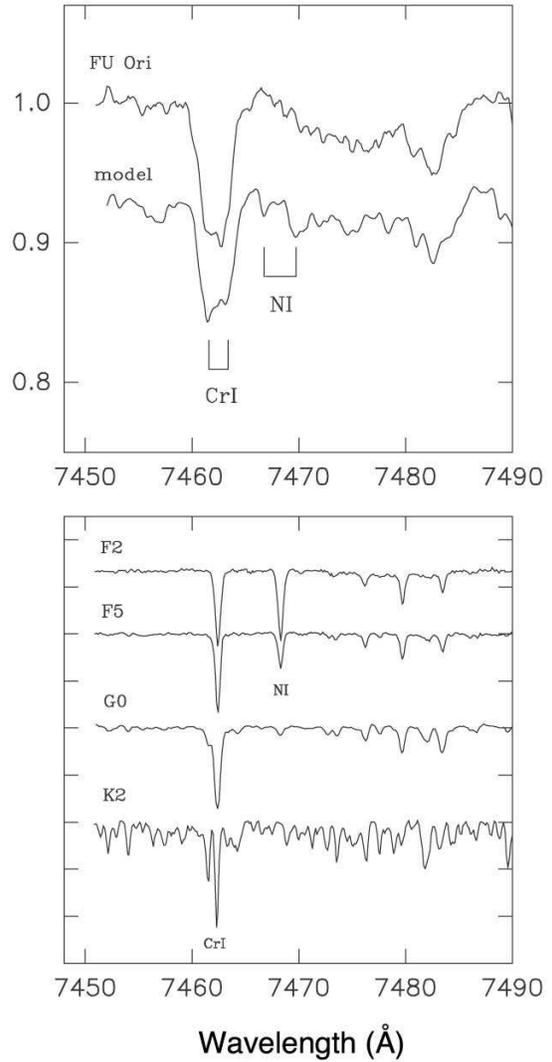}
\caption{  As Fig.\ 17: the line of \ion{N}{1} (EP = 7 eV) which should be split
by the fast rotation of the inner disk is not present in the observed spectrum 
of FU Ori. 
     \label{fig19} }
\end{figure}
%\clearpage

\begin{figure}
%\epsscale{0.75}
\plotone{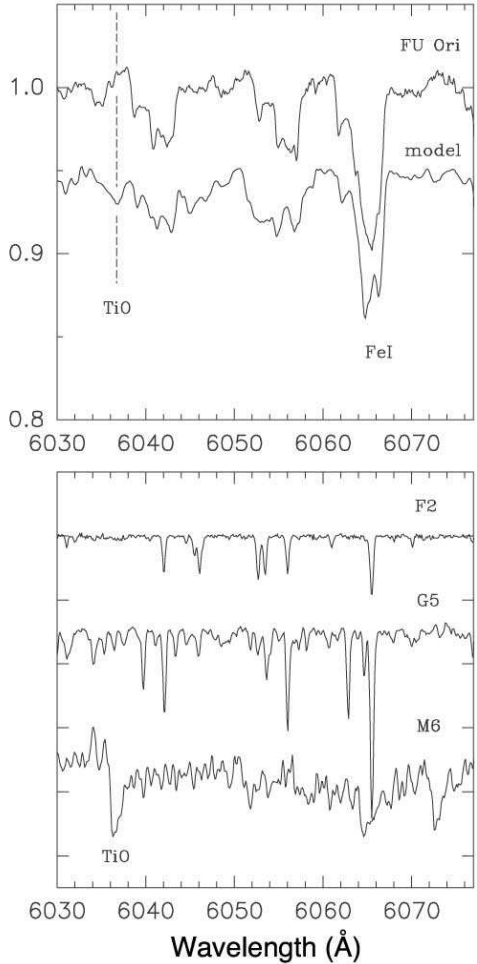}
\caption{ As Fig.\ 17: the ScO/TiO feature predicted by the accretion disk model
is absent in the observed spectrum of FU Ori.
	\label{fig20} }
\end{figure}

%\clearpage

\begin{figure}
%\epsscale{1.0}
\plotone{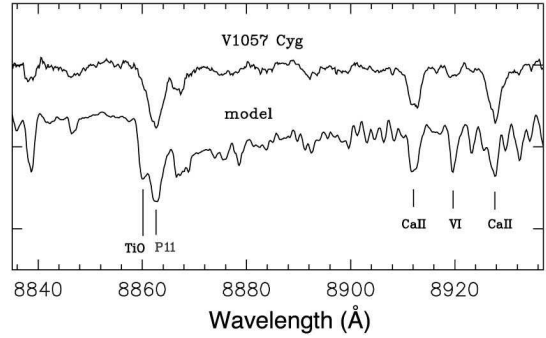}
\caption{ The same as Fig.\ 17 but for V1057 Cyg, an average of the SOFIN spectra of 1998,
1999, and 2000 when TiO was not present in the shell. Note the absence of the TiO feature in
the observed stellar spectrum. 
	\label{fig21} }
\end{figure}
%\clearpage

\begin{figure}
%\epsscale{1.0}
\plotone{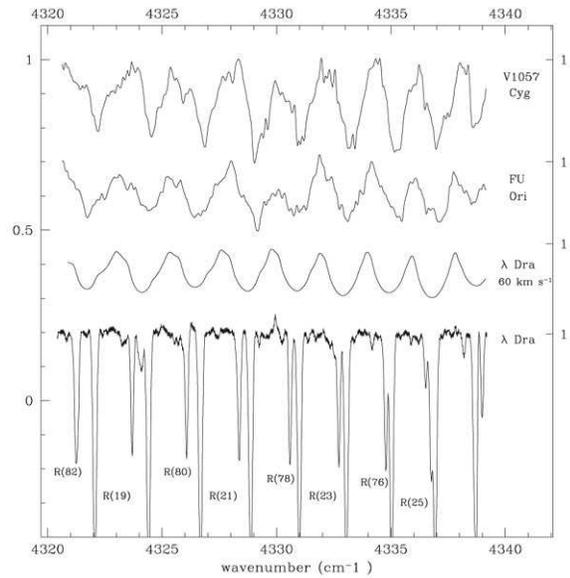}
\caption{At top: the spectra of V1057 Cyg and FU Ori in the 2.3 $\mu$m region,
as observed by K. Hinkle on 1999 October 24.  They have been smoothed by a 5-pixel
box and corrected for their nominal optical-region velocities. 
The FTS spectrum of $\lambda$ Dra (M0\, III) from the atlas of \citet{wal96} has been slightly
smoothed (bottom) and spun up to
$v_{\rm eq}\, \sin i = 60$ km s$^{-1}$ (above).
   \label{fig22}}
\end{figure}
%\clearpage

\begin{figure}
%\epsscale{1.0}
\plotone{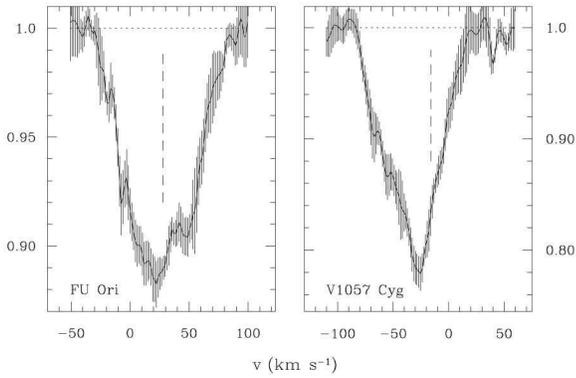}
\caption{Mean CO profiles for V1057 Cyg (average of 5 lines) and FU Ori
(4 lines).  The vertical solid lines show the $\pm$1 standard deviations of
each mean point.  The vertical dashed line indicates the optical velocity
of each star.
   \label{fig23}}
\end{figure}

%\clearpage

\begin{figure}
%\epsscale{1.0}
\plotone{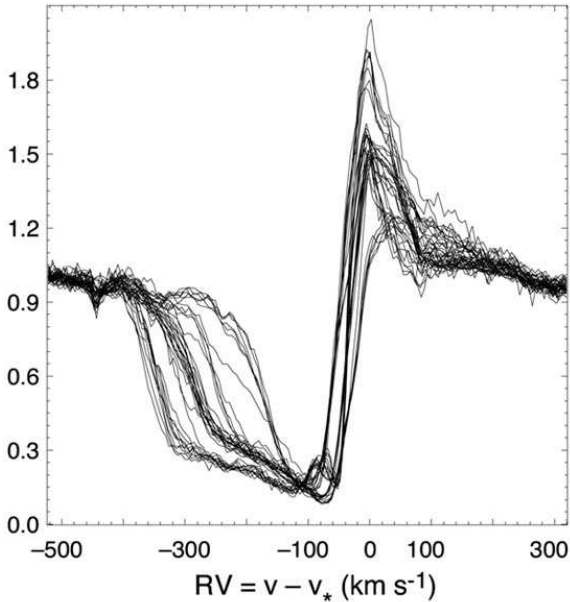}
\caption{Variations of the H$\alpha$ profile of V1057 Cyg in 1996--2001.
	\label{fig24} }
\end{figure}
%\clearpage

\begin{figure}
%\epsscale{1.0}
\plotone{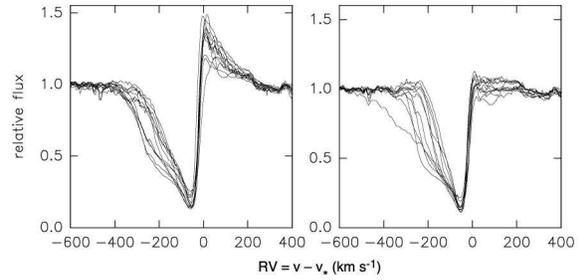}
\caption{Variations of the H$\alpha$ profiles of FU Ori in 1995--2000.
Left panel: all profiles having the emission peak intensity (in continuum
units) greater than 1.20.  Right panel: all those having peak intensity
less than 1.10.  These illustrate how the longward emission peak is
suppressed to different degrees on different occasions.
The extension of the P Cyg absorption wing to different nehative
velocities is also apparent.  There is no obvious correlation between
the two.
	\label{fig25} }
\end{figure}
%\clearpage

\begin{figure}
%\epsscale{0.8}
\plotone{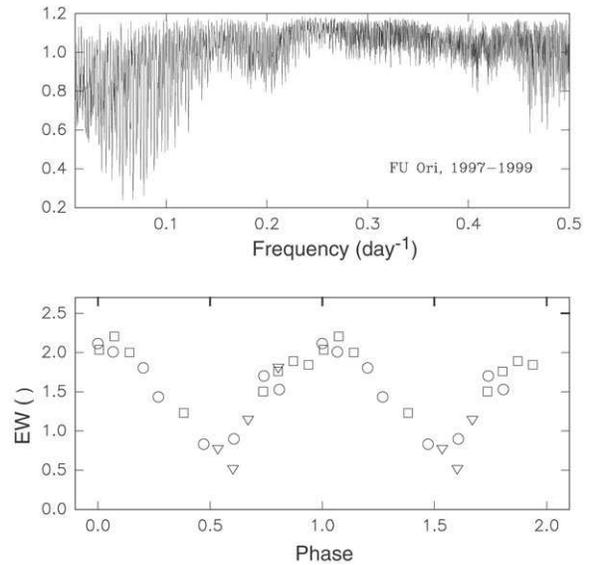}
\caption{Upper panel: periodogram of the variation of EW of the ``fast
wind" section of H$\alpha$ in FU Ori for a period range of 2 to 100 days.
Lower panel: phase diagram for a period of 14.847 days.
Circles: data of 1997; squares: data of 1998; triangles: data of 1999.  
	\label{fig26} }
\end{figure}
%\clearpage

\begin{figure}
%\epsscale{0.7}
\plotone{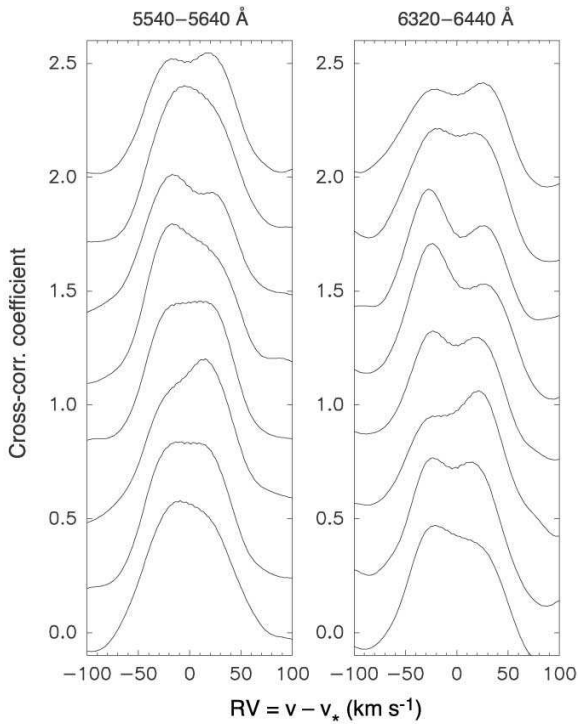}
\caption{Cross-correlation functions, showing night-to-night variability
in the photospheric line profiles in V1057 Cyg over 8 consecutive
nights of August 1997. 
	\label{fig27} }
\end{figure}
%\clearpage

\begin{figure}
%\epsscale{0.7}
\plotone{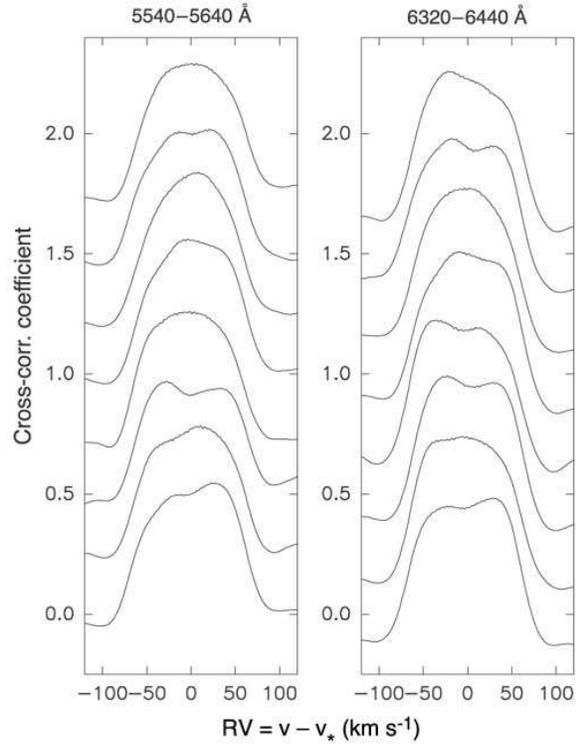}
\caption{Cross-correlation functions, showing night-to-night variability
in the photospheric line profiles in FU Ori in December 1997. 
	\label{fig28} }
\end{figure}
%\clearpage

\begin{figure}
%\epsscale{1.0}
\plotone{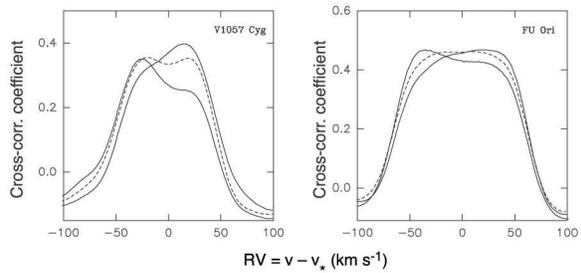}
\caption{Three cross-correlation functions, showing the typical
variations of the photospheric line profiles in V1057 Cyg and FU Ori:
the blue-shifted, centered, and red-shifted profiles. 
	\label{fig29} }
\end{figure}
%\clearpage

\begin{figure}
\plotone{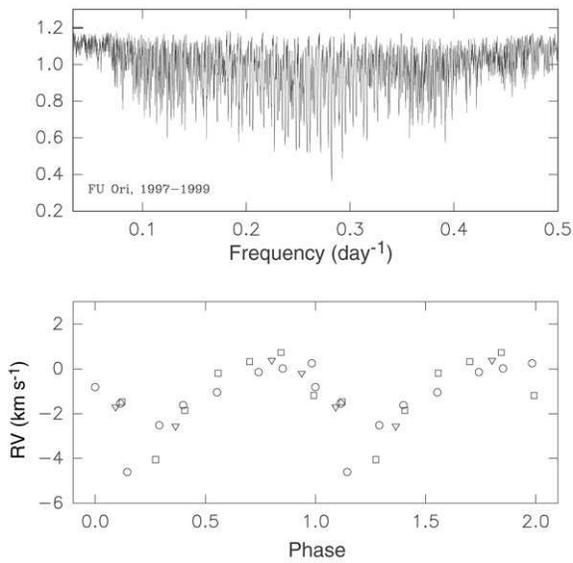}
%\epsscale{0.7}
\caption{Upper panel:  radial velocity periodogram for FU Ori in 1997--1999,
for a period range from 2 to 30 days. The most probable period is
about 3.543 days;
Lower panel: phase diagram for the period of 3.543 days;
circles: data of 1997; squares: data of 1998; triangles: data of 1999. 
	\label{fig30} }
\end{figure}
%\clearpage

\begin{figure}
\plotone{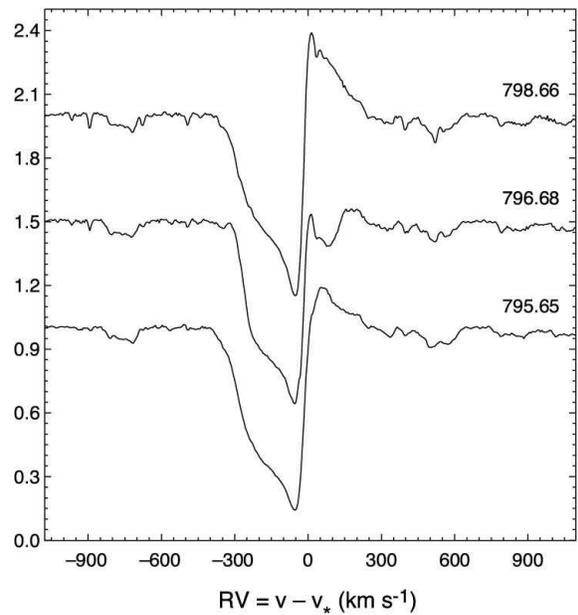}
\caption{Three SOFIN spectra of the H$\alpha$ region of FU Ori, obtained
on the dates indicated (which are JD - 2450000.)  The absorption
component at about RV = +90 km s$^{-1}$ in the longward wing of H$\alpha$
was present on the middle date, but not on the day before or two days later.
        \label{fig31}}
\end{figure}
%\clearpage

\begin{figure}
%\epsscale{1.0}
\plotone{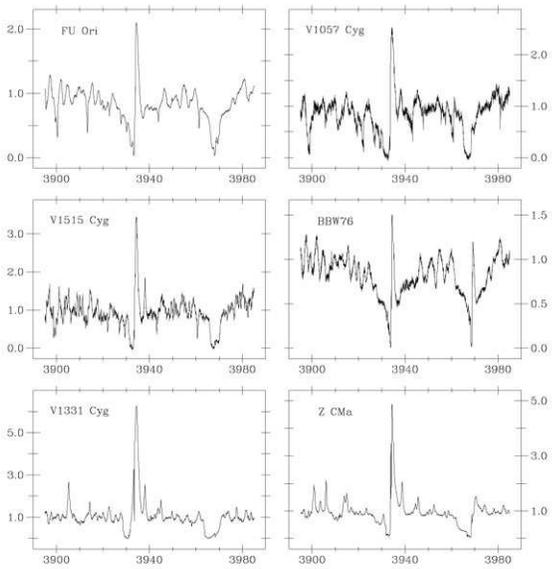}
\caption{The region of the \ion{Ca}{2} H and K lines in six stars, among them
three classical FUors, as observed with HIRES on 1998 October 30, 31.  The
narrow emission line closely longward of $\lambda$3933 in several objects is
\ion{Fe}{2} $\lambda$3938.29 [RMT 3].
        \label{fig32}}
\end{figure}
%\clearpage

\clearpage

\begin{deluxetable}{llllclll}
%% table 22.1.txt = Table 1
\tablecolumns{8}
\tablewidth{0pc}
\tablecaption{H$\alpha$ Structure in V1057 Cyg}
\tablehead{
\colhead{} &\multicolumn{1}{c}{} & \multicolumn{2}{c}{Absorption} &
\colhead{} &\multicolumn{2}{c}{Emission} & \multicolumn{1}{c}{} \\
\cline{3-4} \cline{6-7} \\
\colhead{Source} & \colhead{Date}   & \colhead{EW (\AA)} & \colhead{A$_{c}$} & 
\colhead{} &\colhead{EW (\AA)} & \colhead{E$_{c}$}   & \colhead{$V$ mag.}  } 

\startdata
BM    & 1981 Jun 23 & 1.86      & 0.41, 0.40 &\phn & 0.77 & 0.19 & 11.2 \\
BM    & 1982 May 15  & 0.24+0.24 & 0.15, 0.21 & & 1.00 & 0.27 & 11.35 \\
BM    & 1982 Jul 12 & 1.36      & 0.55, 0.13 & & 0.99 & 0.22 & 11.35 \\
BM    & 1982 Jul 13 & 1.40      & 0.55, 0.14 & & 0.95 & 0.22:& 11.35 \\
C     & 1984 Oct 10 & 2.24+0.84 & 0.59, 0.56 & & 0.96 & 0.35 & 11.5 \\
C     & 1984 Dec 6  & 3.44      & 0.52       & & 0.91 & 0.58:& 11.5  \\
Lick  & 1985 Sep 23 & 4.2       & 0.66       & & 0.62 & 0.28 & 11.6  \\   
C     & 1985 Nov 24 & 0.14+1.60 & 0.32, 0.55 & & 1.79 & 0.93 & 11.6  \\
Budge & 1988 Aug 25 & 7.3       & 0.69       & & 0.65 & 0.34 & 11.6  \\
Welty & 1988 Nov 30 & 7.6       & 0.70      & & 2.2  & 0.92 & 11.6 \\
HC    & 1992 Dec 7  & 5.8       & 0.90      & & \nodata& 0.04& 11.6 \\
SOFIN & 1996 Oct 30 & 3.52      & 0.89       & & 0.96 & 0.45 & 12.35 \\
HIRES & 1997 Aug 13 & 4.24      & 0.87       & & 1.03 & 0.51 & 12.45 \\
SOFIN & 1997 Aug 15--22 & 4.28    & 0.88       & & 1.13 & 0.56 & 12.45 \\
SOFIN & 1998 July 2-15 & 4.01    & 0.84       & & 1.66 & 0.83 & 12.95 \\
SOFIN & 1999 Jul 23--Aug 3 & 5.31 & 0.86      & & 0.60 & 0.52 & 12.82 \\
SOFIN & 2000 Aug7--10 & 3.57     & 0.86       & & 1.23 & 0.51 & 12.59 \\
SOFIN & 2001 Jul 29--Aug 9 & 3.26 & 0.92      & & 0.44 & 0.22 & 12.5 \\
HIRES & 2002 Dec 16 & 3.06      & 0.84       & & 1.71 & 0.62 & 12.5 \\
 
\enddata

{\tablecomments { 
\noindent
Col.1: the source abbreviations are: \\
\hspace*{0.5in} BM: \citealt{bas85} \\
\hspace*{0.5in} C : \citealt{cro87} \\
\hspace*{0.5in} Lick, unpublished; low resolution ($\approx$ 5200) \\
\hspace*{0.5in} Budge, unpublished; Palomar 60-inch, echelle \\
\hspace*{0.5in} \citealt{wlt92} \\
\hspace*{0.5in} HC: \citealt{har95} \\
\hspace*{0.5in} SOFIN: Nordic Optical Telescope, this paper \\
\hspace*{0.5in} HIRES: Keck I, this paper \\
Cols. 3 and 5: EW in the equivalent width of the feature \\
Col.5: A$_{c}$ is here the central depth of the deepest point in
the absorption line, measured downward from the continuum, in continuum
units \\
Col.7: E$_{c}$ is here the central height of the emission peak,
measured upward from the continuum, in continuum units.  It is
subject to rapid night-to-night variations: see \S 4.4.1 \\  } }

\end{deluxetable}

\begin{deluxetable}{cclcccclcccclcc}
%% table23.1g.txt = Table 2
\tabletypesize{\scriptsize}
\tablecolumns{15}
\tablewidth{0pc}
\tablecaption{Shell Line Structure in V1057 Cyg}
\tablehead{
\colhead{} & \colhead{$v$ } & \colhead{$\tau_{0}$} &
\colhead{$\sigma$} & \colhead{EW } & \colhead{} &
\colhead{$v$ } & \colhead{$\tau_{0}$} &
\colhead{$\sigma$} & \colhead{EW } & \colhead{} &
\colhead{$v$ } & \colhead{$\tau_{0}$} &
\colhead{$\sigma$} & \colhead{EW } \\
\colhead{} & \colhead{(km s$^{-1}$)} & \colhead{} &
\colhead{ (km s$^{-1}$)} & \colhead{ (m\AA)} & \colhead{} &
\colhead{ (km s$^{-1}$)} & \colhead{} &
\colhead{ (km s$^{-1}$)} & \colhead{ (m\AA)} & \colhead{} &
\colhead{ (km s$^{-1}$)} & \colhead{} &
\colhead{ (km s$^{-1}$)} & \colhead{ (m\AA)} }

\startdata
  &\multicolumn{4}{c}{Ti I [5] 4997.099} &   &
             \multicolumn{4}{c}{Ti I [38] 4999.504} &  &
             \multicolumn{4}{c}{Ba II [2] 6496.896}  \\  [3pt]
%\\ [-5pt]
\cline{2-5} \cline{7-10} \cline{12-15}\\[3pt]
1  & \nodata & \nodata & \nodata &  \nodata &\phn
    & $-$128.\phn & 0.08  & 3.9 & \phn13.  &\phn
    & $-$128.\phn & 0.23  & 8.3 & \phn96.   \\
2  & $-$107.\phn & 0.022: & 5.4: & \phn5.:  &\phn
    & $-$107.5 & 0.27  & 5.1 & \phn53.  &\phn
    & $-$107.\phn & 0.27  & 6.5 & \phn86.   \\
3  & $-$90. & 0.26  & 3.6 & 36.  &\phn
    & \phn$-$88.5 & 0.87  & 6.0 & 165. &\phn
    & \phn$-$92.5 & 0.33  & 5.5 & \phn89.   \\
4  & $-$78. & 0.33  & 4.5 & 56.  &\phn
    & $-$78. & 1.17  & 6.0 & 202. &\phn
    & $-$80. & 0.82  & 7.8 & 267.  \\
5  & $-$59.5& 0.27  & 4.8 & 49. &\phn
    & \phn$-$61.5 & 1.77  & 5.1 & 221. &\phn
    & $-$61. & 1.06  & 6.5 & 265.  \\
\\
&\multicolumn{4}{c}{Li I [1] 6707.81 } &   &
             \multicolumn{4}{c}{Rb I [1] 7800.227} &  &
             \multicolumn{4}{c}{Rb I [1] 7947.60 }  \\
\cline{2-5} \cline{7-10} \cline{12-15}\\
1  & $-$128.\phn & 0.17  & 8.0 & \phn72.  &\phn
    & $-$129.\phn & 0.075  & 6.9 & \phn33.  &\phn
    & $-$126.\phn & 0.078  & 4.9 & \phn25.   \\
2  & $-$104.\phn & 0.31  & 8.7 & 136.  &\phn
    & $-$106.\phn & 0.085 & 7.7 & \phn41.  &\phn
    & $-$100.\phn & 0.065 & 4.9 & \phn21.   \\
3  & \nodata & \nodata  & \nodata & \nodata  &\phn
    & \nodata & \nodata  & \nodata & \nodata  &\phn
    & \nodata & \nodata  & \nodata & \nodata   \\
4  & $-$82. & 0.96  & 8.3 & 326.  &\phn
    & \phn$-$79.5& 0.32  &10.\phn\phn  & 194.  &\phn
    & $-$81. & 0.21  & 9.8 & 127.  \\
5  & \phn$-$60.5& 1.29  & 8.0 & 389. &\phn
    & \phn$-$58.5& 0.52  & 4.6 & 131. &\phn
    & $-$60. & 0.32  & 5.3 & 101.  \\

\enddata

{\tablecomments { 
\noindent
The symbols are:  $v$ is the central (heliocentric) velocity of that component;
$\tau_{0}$ is its central optical thickness and $\sigma$ is
its width, both expressed as gaussian parameters as explained in the text; EW
is its equivalent width.  \\  } }

\end{deluxetable}

\begin{deluxetable}{cccc}
%%  table23.1.txt = Table 3
\tablecolumns{4}
\tablewidth{0pc}
\tablecaption{\ion{Ti}{1} Shell Line Structure, 1997 Aug. 12, 13 }
\tablehead{
\colhead{Component} & \colhead{$\xi_{0}$} & \colhead{T$_{\rm exc}$} &
 \colhead{log N(\ion{Ti}{1})} \\
\colhead{(km s$^{-1}$) } & \colhead{(km s$^{-1}$) } &\colhead{($^{\circ}$K}) &
\colhead{(cm$^{-2}$) } }

\startdata

$-$128  & \phd1.3:  & \phd4000.: & \phd13.38: \\
$-$107  & 1.9   & 4350.  & 14.08  \\
\phn$-$89  & 3.1   & 3650.  & 14.78  \\
\phn$-$78  & 4.7   & 3600.  & 15.05  \\
\phn$-$61  & 4.8   & 3700.  & 14.95  \\ 
 
\enddata
 
\end{deluxetable}

\begin{deluxetable}{llllclll}
%% tablex.txt = Table 4
\tablecolumns{7}
\tablewidth{0pc}
\tablecaption{Equivalent Widths (in \AA) of Emission Cores }
\tablehead{
\colhead{} & \colhead{}   & \colhead{\ion{Fe}{1}} & \colhead{\ion{Fe}{1}} & 
 \colhead{\ion{Fe}{1}} & \colhead{\ion{Fe}{1}}   & \colhead{\ion{Fe}{2}} \\  
\colhead{Date} & \colhead{$V$}   & \colhead{5615} & \colhead{6191} & 
 \colhead{6393} & \colhead{6400}   & \colhead{6516}  } 
\startdata
1996 Oct & 12.35 & 0.154 &\nodata & 0.156 & 0.236 & 0.210 \\
1997 Aug & 12.45 & 0.237 & 0.310  & 0.220 & 0.319 & 0.266 \\
1998 Jul & 12.95 & 0.205 & 0.306  & 0.175 & 0.268 & 0.226 \\
1999 Jul & 12.82 & 0.179 & 0.285  & 0.190 & 0.246 & 0.180 \\
2000 Aug & 12.58 & 0.192 & 0.291  & 0.170 & 0.256 & 0.239 \\
2001 Aug & 12.48 & 0.260 & 0.240  & 0.200 & 0.271 & 0.245 \\
\enddata
\end{deluxetable}

\begin{deluxetable}{ll}
%% table3.1.txt =  Table 5
\tablecolumns{2}
\tablewidth{0pc}
\tablecaption{Template Stars}
\tablehead{
\colhead{Star} & \colhead{MK Type} } 
\startdata
32 Aql & F2 Ib \\
41 Cyg & F5 II \\
$\beta$ Aqr & G0 Ib \\
9 Peg  & G5 Ib \\
40 Peg & G8 II \\
43 Tau & K2 III \\
$\beta$ And & M0 IIIa \\
72 Leo & M3 Iab-Ib \\
30 Her & M6 IIIa \\
 
\enddata
\end{deluxetable}

\begin{deluxetable}{llll}
%% table3.2.txt = Table 6 
\tablecolumns{4}
\tablewidth{0pc}
\tablecaption{Flux Contributions to Disk}
\tablehead{
\colhead{Type} & \colhead{5500 \AA\ } & \colhead{6400 \AA\ }
               & \colhead{9000 \AA\ } } 
\startdata
F2-F7: & 0.5 & 0.4 & 0.3 \\
F8-K2: & 0.4 & 0.4 & 0.35 \\
K5-M6: & 0.1 & 0.2 & 0.35 \\
\cline{2-4} \\
total: & 1.0 & 1.0 & 1.0 \\ 
\enddata

\end{deluxetable}

\begin{deluxetable}{lcccc}
%% Table 7
\tablecolumns{5}
\tablewidth{0pc}
\tablecaption{Fits to the CO Lines}
\tablehead{
\colhead{} & \colhead{} &
\colhead{$v_{\rm eq}\, \sin i$} &\colhead{$v$} & \colhead{A$_{c}$} \\
\colhead{Star} & \colhead{Component}   & \colhead{(km s$^{-1}$)} &
\colhead{(km s$^{-1}$)}
 & \colhead{}   } 
\startdata
V1057 Cyg &  1  &  12.  &  $-$26.  &  0.085 \\
          &  2  &  45.  &  $-$35.  &  0.15\phn  \\
FU Ori    &  1  &  48.  &  +27.  &  0.105 \\

\enddata
{\tablecomments { 
\noindent
A$_{c}$ is defined in Table 1.
  \\  } }

\end{deluxetable}

\begin{deluxetable}{cccc}
%% petrov49.wind_ew.dat = table8
\tablecolumns{4}
\tablewidth{0pc}
\tablecaption{EWs of H$\alpha$ P Cyg absorption in FU Ori }
\tablehead{
  \colhead{JD-2450000 } & \colhead{EW (\AA ) } &
  \colhead{JD-2450000 } & \colhead{EW (\AA ) } }

\startdata

0054.539  &  1.478  & 1092.721  &  2.035 \\
0055.575  &  1.177  & 1093.704  &  2.206 \\
0795.647  &  2.114  & 1094.713  &  1.999 \\
0796.676  &  2.007  & 1471.693  &  0.774 \\
0798.665  &  1.805  & 1472.722  &  0.516 \\
0799.706  &  1.432  & 1473.689  &  1.144 \\
0802.678  &  0.831  & 1475.717  &  1.808 \\
0804.697  &  0.897  & 1887.569  &  1.231 \\
0806.678  &  1.701  & 1888.527  &  1.133 \\
0807.695  &  1.530  & 1889.571  &  1.745 \\
0890.399  &  1.229  & 1890.572  &  1.000 \\
1088.672  &  1.504  & 1892.573  &  1.308 \\
1089.706  &  1.759  & 1893.568  &  1.308 \\
1090.704  &  1.890  & 1896.572  &  0.755 \\
1091.706  &  1.843  &\nodata    & \nodata \\
\enddata
\tablecomments {
 The equivalent widths are for the section of the absorption line
 between RV = $-$110 and $-$270 km s$^{-1}$  }

\end{deluxetable}

\begin{deluxetable}{rrcrr}
%% table 34.1.txt = Table 9
\tablecolumns{4}
\tablewidth{0pc}
\tablecaption{Radial Velocities from Cross-Correlation Functions}
\tablehead{ \multicolumn{2}{c}{V1057 Cyg} & &
 \multicolumn{2}{c}{FU Ori} \\
\cline{1-2} \cline{4-5} \\
\colhead{JD-2450000} & \colhead{RV (km s$^{-1}$)} &
  & \colhead{JD-2450000} & \colhead{RV (km s$^{-1}$)} }
\startdata
   387.448 &  $-$1.590 & &      54.538 &   1.139 \\
   676.572 &  $-$1.067 &  &     55.575 &   0.461 \\
   677.398 &  $-$2.746 &  &    795.647 &  $-$0.800 \\
   678.402 &   3.270 &  &     796.676 &  $-$2.501 \\
   679.395 &  $-$0.386 &  &     798.665 &   0.029 \\
   680.384 &  $-$4.435 &  &     799.705 &  $-$4.599 \\
   681.384 &  $-$6.608 &  &     802.677 &   0.256 \\
   682.383 &  $-$1.637 &  &     804.697 &  $-$1.040 \\
   683.405 &   1.017 &  &     806.678 &  $-$1.536 \\
   997.472 &   4.266 &  &     807.694 &  $-$1.601 \\
   999.601 & $-$10.106 &  &     890.398 &  $-$0.140 \\
  1000.562 &  $-$1.914 &  &    1088.672 &   0.330 \\
  1002.532 &   0.967 &  &    1089.706 &  $-$1.186 \\
  1003.546 &  $-$8.107 &  &    1090.703 &  $$-$$4.049 \\
  1005.521 &  $-$5.065 &  &    1091.706 &  $-$0.190 \\
  1008.545 &  $-$0.917 &  &    1092.721 &   0.730 \\
  1383.554 &   1.829 &  &    1093.704 &  $-$1.470 \\
  1384.603 &  $-$2.308 &  &    1094.713 &  $-$1.849 \\
  1385.556 & $-$10.181 &  &    1471.693 &   0.371 \\
  1386.550 &   5.770 &  &    1472.722 &  $-$1.709 \\
  1387.515 &   0.604 &  &    1473.689 &  $-$2.569 \\
  1388.529 &   1.606 &  &    1475.717 &  $-$0.206 \\
  1389.522 &  $-$8.802 &  &    1887.569 &  $-$0.369 \\
  1390.555 &  $-$0.839 &  &    1888.527 &  $-$1.780 \\
  1391.524 &   0.689 &  &    1889.571 &   1.290 \\
  1392.552 &   2.195 &  &    1890.572 &   3.721 \\
  1393.586 &  $-$3.338 &  &    1892.573 &   1.160 \\
  1394.561 &  $-$4.708 &  &    1893.568 &  $-$0.624 \\
  1764.606 &   4.634 &  &    1896.572 &  $-$1.344 \\
  1765.611 &   2.115 &  & \nodata & \nodata \\
  1766.599 &   3.804 &  & \nodata & \nodata \\
  1767.609 &   9.921 &  & \nodata & \nodata \\
  2120.480 &  $-$0.997 &  & \nodata & \nodata \\
  2121.493 &   1.509 &  & \nodata & \nodata \\
  2123.571 &   3.578 &  & \nodata & \nodata \\
  2124.495 &  $-$2.060 &  & \nodata & \nodata \\
  2125.540 &  $-$2.953 &  & \nodata & \nodata \\
  2127.595 &   2.997 &  & \nodata & \nodata \\
  2128.557 &  $-$3.084 &  & \nodata & \nodata \\
  2129.584 &   0.271 &  & \nodata & \nodata \\
  2130.568 &  $-$1.492 &  & \nodata & \nodata \\
  2131.501 &  $-$1.674 &  & \nodata & \nodata \\
\enddata
{\tablecomments {RV is the measured radial velocity minus the
radial velocity of the star: $-$16 km s$^{-1}$ for V1057 Cyg and +28
km s$^{-1}$ for FU Ori.}}

\end{deluxetable}

\end{document}